\documentclass[11pt]{article}
\usepackage[T1]{fontenc}

\usepackage[margin=1in]{geometry}

\usepackage[hidelinks]{hyperref}
\hypersetup{pdfauthor={},pdftitle={List Update with Time Windows and Delays without Batch-Serving}}
\usepackage{graphicx}
\usepackage{xcolor}
\usepackage{bbm}
\usepackage{algorithm}
\usepackage{algpseudocode}
\usepackage{amsmath}
\usepackage{amsthm}
\usepackage{array}
\usepackage{amssymb}
\usepackage{makecell}
\usepackage{threeparttable}
\usepackage{thmtools}
\usepackage{array}
\usepackage{authblk}
\newcolumntype{C}[1]{>{\centering\arraybackslash}m{#1}}

\usepackage[capitalise, noabbrev]{cleveref}
\makeatletter
\renewcommand{\theHALG@line}{\thealgorithm.\arabic{ALG@line}}
\makeatother

\theoremstyle{plain}
\newtheorem{theorem}{Theorem}[section]
\newtheorem{lemma}[theorem]{Lemma}

\newtheorem{corollary}[theorem]{Corollary}

\theoremstyle{definition}

\theoremstyle{remark}
\newtheorem{remark}[theorem]{Remark}

\definecolor{ForestGreen}{rgb}{0.1333,0.5451,0.1333}
\definecolor{DarkRed}{rgb}{0.8,0,0}
\definecolor{Red}{rgb}{1,0,0}

\pdfoutput=1

\title{List Update with Time Windows and Delays without Batch-Serving\footnote{Research partly supported  by NSF grant CCF-2218677, ONR grant ONR-13533312,
and the Wai T. Chang Chair in Industrial Engineering and Operations Research.Authors emails: \tt{icey.ai@columbia.edu}, \tt{jl6639@columbia.edu},  and \tt{cliff@ieor.columbia.edu}.}} 

\author[1]{Icey Siyi Ai}
\author[1]{Jingwei Li}
\author[1]{Clifford Stein}

\affil[1]{Department of Industrial Engineering and Operations Research,
Columbia University.  }

\usepackage{etoolbox}
\begin{document}

\maketitle

    \algdef{SE}[SUBALG]{Indent}{EndIndent}{}{\algorithmicend\ }
    \algtext*{Indent}
    \algtext*{EndIndent}

\newcommand{\eps}{\varepsilon}
\newcommand{\Bern}{\mathrm{Bern}}

\newcommand{\calD}{\mathcal{D}}
\newcommand{\calN}{\mathcal{N}}
\newcommand{\calH}{\mathcal{H}}

\newcommand{\RR}{\mathbb{R}}
\newcommand{\EE}{\mathbb{E}}
\newcommand{\PP}{\mathbb{P}}
\newcommand{\NN}{\mathbb{N}}
\newcommand{\QQ}{\mathbb{Q}}
\newcommand{\FF}{\mathbb{F}}
\newcommand{\calA}{\mathcal{A}}
\newcommand{\calE}{\mathcal{E}}
\newcommand{\calS}{\mathcal{A}}
\newcommand{\remove}{\char`\\}
\newcommand{\parenth}[1]{\left( #1\right)}

\newcommand{\bfx}{\mathbf{x}}

\newcommand{\lrangle}[1]{\left< #1 \right>}
\newcommand{\lrparenth}[1]{\left( #1 \right)}
\newcommand{\lrbracket}[1]{\left[ #1 \right]}

\newcommand{\ind}[1]{\mathbbm{1}\left\{ #1 \right\}}

\newif\ifcomment
\commentfalse

\ifcomment
    \newcommand{\jingwei}[1]{{\color{violet} Jingwei: #1}}
    \newcommand{\cliff}[1]{{\color{red} Cliff: #1}}
    \newcommand{\icey}[1]{{\color{orange} Icey: #1}}
    \newcommand{\old}[1]{{\color{lightgray} #1}}
\else
    \newcommand{\jingwei}[1]{}
    \newcommand{\cliff}[1]{}
    \newcommand{\icey}[1]{}
    \newcommand{\old}[1]{}
\fi

\newcommand{\OPT}{\textnormal{\textsc{OPT}}}
\newcommand{\ALG}{\textsc{ALG}}
\newcommand{\FER}{\textnormal{\textsc{FER}}}
\newcommand{\WBIT}{\textnormal{\textsc{WBIT}}}
\newcommand{\ALGDELAY}{\textnormal{\textsc{DMF}}}
\newcommand{\pos}{\mathrm{pos}}
\newcommand{\bit}{\textit{bit}}
\newcommand{\E}{\mathbb{E}}

\newcommand{\pF}{\prec_{\FER}}          
\newcommand{\pO}{\prec_{\OPT}}          
\newcommand{\tO}[1]{t^{\OPT}_{#1}}      
\newcommand{\tF}[1]{t^{\FER}_{#1}}      
\newcommand{\pW}{\prec_{\WBIT}}

\thispagestyle{empty}

\begin{abstract}
We study list update with time windows and delays when every request requires a separate access. Unlike batch-serving models, serving several requests to one element still incurs one access per request, and each access may change the list for subsequent requests. For time windows of overlap at most $k\ge2$, our deterministic First-Element Removal algorithm is $(3-2/k)$-competitive, hence $3$-competitive in general. Against an oblivious adversary, a randomized adaptation of BIT is $7/4$-competitive on arbitrary $2$-overlap inputs. More generally, on inputs without self-overlap, it is $\left(13/4-9/[2(k+1)]\right)$-competitive for every $k\ge2$. For delays, we give a deterministic $5$-competitive algorithm and a lower bound of $2+\sqrt3$ for deterministic delay-unaware algorithms. Our analyses extend inversion potentials with per-request credits that reconcile different service times and orders.
\end{abstract}

\section{Introduction}\label{sec: intro}
The list update problem is a fundamental problem in online algorithms and competitive analysis. First studied by McCabe \cite{mccabe1965serial} as a model for maintaining a sequential file, it was later used by Sleator and Tarjan \cite{MTF} to introduce competitive analysis. 
In the classical list update problem, requests for elements in a list arrive online and must be accessed and served immediately. Serving an element costs its current position in the list; immediately after the access, the algorithm may move the accessed element forward for free, and it may also swap two adjacent elements at a cost of 1 at any time. The objective is to minimize the total access and paid-swap cost.

Many online settings relax the requirement that requests must be served immediately upon arrival. Two common relaxations are {\em time windows}, where a request may be served any time between its release time and deadline, and  {\em delays}, where a request may be served later at the cost of accumulating a delay penalty. Such relaxations have been studied for various online problems, including paging (caching)\cite{pagingTW}, multilevel aggregation\cite{aggDelay,CarrascoPSV18}, and set cover\cite{setcoverdelay}. They capture situations in which an online algorithm can wait in order to reorder, batch, or otherwise coordinate its actions. 

In recent work, Azar et al.  \cite{LUTWazar} introduced time window and delay models for the list update problem. A distinctive feature of their model is a \emph{batch-serving} rule: a single access may serve any set of alive requests, at a cost equal to the position of the furthest element in the set. This rule is powerful and natural in settings where one action can satisfy many pending requests, but it can be too strong when requests represent distinct operations that must each be executed. For example, consider an adaptive data service that keeps records or cached objects in a priority list. The position of an object represents the cost of locating or loading it, and accessing the object may justify promoting it in the list. Each request, however, is a separate transaction on a specified object, possibly with its own user, input, validation, or update. Thus, even if several requests are handled in the same time step, each request incurs its own access cost. 




Motivated by such applications, in this paper we study the list update problem with time windows and delays, {\em without batch serving}: each request must still be paid for separately, even if several requests are served at the same time.  Because each access may trigger a free forward move that reconfigures the list, the order in which simultaneous or overlapping requests are served becomes part of the algorithmic decision. 
We call this the \textit{No-Batching-Allowed list update} model with time windows or delays, or \textit{NBA list update} for short.
It captures settings where each request is a distinct unit of work, so that serving one request does not automatically serve the others. 
The resulting model is incomparable with the batch-serving model of Azar et al.\cite{LUTWazar}: both generalize the classical list update problem in \cite{mccabe1965serial}, but neither contains the other. As we will see below, our competitive ratios are significantly smaller than those in \cite{LUTWazar}, but a direct comparison is not particularly informative.
Removing batch-serving introduces a new difficulty when time windows overlap: the algorithm must choose an order in which to serve pending requests, and each access may rearrange the list for the remaining requests. We capture this with an overlap parameter $k$: an instance is \emph{$k$-overlap} if at most $k$ time windows contain any given time $t$, including time windows of requests for the same list item. When $k=1$, the problem reduces to the classical setting. This parameter has a natural operational interpretation as a bound on simultaneous demand, such as the number of open sessions, active users, queued transactions, or pending tasks a system may have at once. Small $k$ corresponds to lightly loaded or tightly scheduled systems, while large
$k$ captures higher congestion.
Besides bounds for general inputs, we obtain refined guarantees in terms of the maximum overlap. For example, our deterministic algorithm \FER{} is 3-competitive under general inputs, but also $(3-2/k)$-competitive for $k$-overlap, interpolating between the general guarantee of $3$ and the competitive lower bound $2$ at $k=2$, achieving optimum under the latter case.


Azar et al.~\cite{LUTWazar} leave open whether randomization can improve their deterministic bounds for the batch-serving models. We obtain such improvements in our non-batch time-window model on 
inputs \emph{without self-overlap}, meaning that the time windows of requests to the same element are disjoint. This condition has a natural operational interpretation in systems where clients interact sequentially with a shared server. Each list element may store information associated with one client, who waits for a response before submitting the next operation. When responses are returned at the requests' deadlines, successive requests from that same client have disjoint windows, although requests from different clients may overlap. On $k$-overlap inputs without self-overlap, our randomized algorithm \textsc{WBIT} achieves competitive ratio $13/4-9/[2(k+1)]$ for every $k\ge2$, improving \textsc{FER}'s guarantee for $2\le k\le7$ and matching it at $k=8$. (FER and WBIT are defined below.)   For $k=2$, we can remove the no-self-overlap restriction and obtain a $7/4$-competitive guarantee, separating the deterministic and randomized regimes by breaking the deterministic lower bound of $2$ in this setting.

\subsection{Our results}\label{sec: ourResults}
\begin{table}[t]
\centering
\begin{threeparttable}
\begin{tabular}{|C{0.20\linewidth}|C{0.20\linewidth}|C{0.25\linewidth}|C{0.20\linewidth}|}
\hline
\hline
\textbf{Setting} & \textbf{General Bound} & \textbf{$k$-overlap Bound} & \textbf{Lower Bound} \\
\hline
\makecell{Time windows\\ Deterministic}& \makecell{3 \\(\cref{sec:fer-upper})} & \makecell{$3-\frac{2}{k}$ \\(\cref{sec:fer-upper})}&  \makecell{$2$ \cite{MTF}}\\
\hline
\makecell{Time windows\\ Randomized} & \makecell{$3$} & \makecell{$\tfrac{13}{4}-\tfrac{9}{2(k+1)}$\tnote{$\ddagger$}\\ (\cref{sec:rand})} & \makecell{$1.5$ \cite{1.5Teia}} \\
\hline
\makecell{Delays} & 5 (\cref{delay_6_comp}) & \makecell{Not applicable} & \makecell{$2+\sqrt3\approx3.732^\S$\\(\cref{delay-lb})}\\
\hline
\end{tabular}
\begin{tablenotes}
\footnotesize
\item[$\ddagger$] This guarantee assumes inputs without self-overlap.
\item[$\S$] This lower bound applies to deterministic delay-unaware algorithms.
\end{tablenotes}
\end{threeparttable}
\caption{Summary of our main results for the NBA time-window and delay models. Randomized guarantees are against an oblivious adversary.}
\label{tab:our-results}
\end{table}

Our results 
are summarized in \cref{tab:our-results}. 

 
\noindent\textbf{The new models.} We formulate {\em No-Batching-Allowed list update with time windows and delays}. Each request requires a separate access and every access allows moving the accessed element forward for free. Consequently, serving one request can change the list and the cost of serving the remaining requests, so the algorithm must choose both when and in what order to serve them. In contrast, the batch-serving models of Azar et al.~\cite{LUTWazar} allow one access to serve several pending requests at once but charge for all swaps. See further related work in~\cref{app:related-work}.

For the batch-serving models, Azar et al.~\cite{LUTWazar} give competitive ratios of $24$ for time windows and $336$ for delays using deterministic algorithms without advance knowledge of deadlines or future delay costs. These results motivate us to study the role of randomization and advance information in our non-batch models.

\noindent\textbf{Deterministic algorithms for time windows.} Our main deterministic algorithm, First-Element Removal (\FER{}), combines the classical move-to-front (MTF)\footnote{In classical list update, the move-to-front (MTF) algorithm serves each request immediately and then moves the accessed element to the front for free. MTF is $2$-competitive~\cite{MTF}, matching the lower bound of $2$ for deterministic online algorithms uniformly over all list sizes~\cite{counterBIT}. } with a simple rule for early service. Choosing this rule is crucial: serving every request upon arrival, at its deadline, or after clearing all cheaper pending requests as in batch-serving can each lead to a competitive ratio of $\Omega(n)$ (\cref{sec:simple-orders}). \FER{} only serves requests to the element at the head of its list early, yet achieves a competitive ratio of $3-\frac2k$ on $k$-overlap inputs for $k\ge2$ (\cref{sec:fer-upper}). 

\noindent\textbf{Randomized algorithms for time windows.} 
Our randomized algorithm \WBIT{} serves requests at deadlines and uses the classical BIT\footnote{BIT assigns each element an independent uniformly random initial bit. It serves each request immediately, moves the accessed element to the front if its bit is $0$, and then flips the bit in either case. BIT is $7/4$-competitive against an oblivious adversary~\cite{counterBIT}.} rule~\cite{counterBIT} to decide whether to move the accessed element to the front. For every $k\ge2$, \WBIT{} is $\left(\frac{13}{4}-\frac{9}{2(k+1)}\right)$-competitive on $k$-overlap inputs without self-overlap (\cref{sec:rand}). This improves \FER{}'s guarantee for $2\le k\le7$ and matches it at $k=8$. For $k=2$, we remove the no-self-overlap restriction and obtain a ratio of $7/4$ on all $2$-overlap inputs, strictly below the deterministic lower bound of $2$.

\noindent\textbf{Algorithm and lower bound for delays.} We distinguish the \emph{delay-aware} regime, in which each request's delay function is known upon arrival, from the \emph{delay-unaware} regime, in which only incurred delay costs are observed. Our Delayed Move-to-Front algorithm (\ALGDELAY{}) uses accumulated delay relative to access cost to determine when to access an element and move it to the front, achieving a competitive ratio of $5$ in both regimes (\cref{sec: delay}). We also prove a lower bound of $2+\sqrt3\approx3.732$ for every deterministic delay-unaware algorithm, separating this setting from classical list update, whose optimal deterministic ratio is $2$. 

\noindent\textbf{Technical contributions.} 
The challenge in analyzing list update with time windows is that \OPT{} and the online algorithm may serve requests at different times.  To address this challenge we need to go beyond the standard inversion counting potential functions.
For each request served first by \OPT{}, the potential retains part of \OPT{}'s access cost and maintains request-specific accounts that track subsequent changes in the two lists. The overlap bound limits how many such accounts a service can affect. For \WBIT{}, these accounts are combined with BIT's bit-weighted inversions~\cite{counterBIT}; for delays, they are applied to groups of requests served consecutively by \ALGDELAY{}.

We also take advantage of the time windows and delays by allowing some request-specific terms in the potential function to be negative until they are served by both $\OPT{}$ and $\ALG$. Once all requests have been served, however, the potential is nonnegative, which is sufficient for the amortized analysis.

Our time-window lower bounds use two constructions. For $h$-local-lazy MTF algorithms, requests are released only after their elements have moved beyond the first $h$ positions, preventing early service. For the tighter \FER{} lower bound, the fixed request order $e_1,\ldots,e_n$ is repeated arbitrarily many times. Repeating this pattern for a fixed list size makes the additive constant negligible. In contrast, the delay lower bound chooses requests adaptively based on the online algorithm's actions, exploiting delay-unawareness.

\subsection{Related work}\label{app:related-work}

\paragraph{Classical algorithms and bounds.}
The survey by Kamali and L\'opez-Ortiz~\cite{kamali2013survey}
provides a comprehensive overview of list-update algorithms
and models. Besides MTF, deterministic $2$-competitive
algorithms include TIMESTAMP~\cite{TIMESTAMPalbers} and the
Move-to-Recent-Item (MRI) family~\cite{mrifamily}.
For randomized algorithms against an oblivious adversary,
the COUNTER family generalizes BIT and gives competitive
ratios arbitrarily close to $\sqrt{3}$~\cite{counterBIT}.
COMB combines BIT and TIMESTAMP and is
$8/5$-competitive~\cite{comb}, while Teia~\cite{1.5Teia}
proved a randomized lower bound of $3/2$.
Further algorithmic results appear
in~\cite{schulz1998two,irani1991two,useNoFreeExchange}.

\paragraph{Alternative cost models.}
Variants of list update include the paid-exchange model, in which all exchanges are paid~\cite{albers2020new}, the partial-cost model~\cite{TIMESTAMPalbers,1.50084AMBUHL20013}, and the uniform-cost model~\cite{basiak_et_al:LIPIcs.ESA.2025.76}.
In the partial-cost model, accessing the element at position
$j$ costs $j-1$, rather than $j$.
For this model, Amb\"uhl et al.~\cite{1.50084AMBUHL20013}
proved a randomized lower bound of approximately $1.50084$. Recent work also studies list update under i.i.d.\ requests,
including near-optimal expected access costs for the
Transposition rule~\cite{coester2026transposition,mizgerd2026transposition}.

\paragraph{Other problems with time windows and delays.}
Time-window and delay models have also been studied in
paging and caching~\cite{azar2017online,pagingTW},
multilevel aggregation~\cite{aggDelay,CarrascoPSV18},
set cover~\cite{setcoverdelay}, matching~\cite{bienkowski2017match,ashlagi2017min,mari2022online,deryckere2022online},
and facility location~\cite{bienkowski2021online}.
Azar and Touitou~\cite{azar2019general} developed a general
framework for online metric optimization problems with
deadlines or delay costs.

\section{Preliminaries}\label{sec: prelim}


\subsection{List update with time windows}

We are given a list $L$ of $n$  distinct elements in a specific initial order. The input is a sequence of $m$ requests $\sigma = \{\sigma_i\}_{i\in[m]}$.
Each request has the form $\sigma_i =(e_i,r_i,d_i)$, consisting of a requested element $e_i\in L$, a release time $r_i$, and a deadline $d_i\geq r_i$. We index requests in nondecreasing order of their release times. The input is online: request $\sigma_i$ is revealed only at time $r_i$.


To serve request $\sigma_i$, an algorithm \ALG{}  must access element $e_i$ at some time in the interval $[r_i,d_i]$. Every request requires a separate access, including requests to the same element.
We use the \textit{full cost model} of Sleator and Tarjan \cite{MTF}: accessing the $j$-th element $e$ of the list costs
$j$; immediately afterward, \ALG{} may move $e$ forward to any position in the list for free. Any adjacent swaps by \ALG{} are allowed at any time with a cost of 1. The objective is to minimize the total access and paid-swap cost. We write $\ALG{}(\sigma)$ for the cost of algorithm \ALG{} on input $\sigma$, and $\OPT{}(\sigma)$ for the optimal offline cost.


We work in integral time. At every time step, all releases are processed before any service actions. Multiple actions may then occur at the same time, but their order must be specified. In particular, each access is immediately followed by its free move before the next action. Thus, when several requests have the same deadline, an algorithm may choose their service order, and each service may change the list before the next request is served.


The window of request $\sigma_i$ contains time $t$ if $r_i\le t\le d_i$. This is a property of the input. For a particular algorithm, $\sigma_i$ is \emph{alive} at time $t$ if its window contains $t$ and the algorithm has not yet served it. We use the parameter $k\geq 1$ to denote \textit{the maximum overlap number} among the time windows of requests. An input is \emph{$k$-overlap} if, for every time $t$, the windows of at most $k$ requests contain $t$. We say that an input has \emph{no self-overlap} if the time windows of any two distinct requests to the same element are disjoint; otherwise, it has \emph{self-overlap}.

The classical list update problem is the special case of this model where every request has a distinct point window, $r_i=d_i$. Thus the time-window model inherits the classical lower bounds, including the deterministic lower bound of $2$ and the randomized lower bound of $1.5$ against an oblivious adversary.


\subsection{List update with delays}

In the delay model, a request $\sigma_i$ specifies an element $e_i$, a release time $r_i$, and a delay function $f_i(t)$ for $t\ge r_i$. If the request is served at time $t$, the algorithm pays $f_i(t)$ in addition to the access and paid-swap costs. As in the time-window model, every request requires a separate access. 

Time in the delay model is continuous. Releases, accesses, and swaps may occur at arbitrary nonnegative real times, and accesses and swaps are instantaneous. At a common time, all releases occur before service actions, and each schedule specifies the order of its accesses and moves.


A request is \emph{pending} if it has been released but has not yet been served. A \emph{delay-aware} algorithm learns $f_i$ when $\sigma_i$ is released.
A \emph{delay-unaware} algorithm continuously observes the current delay of every pending request but does not know future delay values. Further assumptions on the delay functions are given in \cref{sec: delay}.



The delay model also contains classical list update as a special case: delay functions can make postponing a request until the next release extremely expensive. Hence the classical list-update lower bounds apply here as well. 


\subsection{Notation}
For an algorithm \ALG{}, let $\pos^t_{\ALG}(e)$ be the position of element $e$ in \ALG{}'s list at time $t$; we omit $t$ or \ALG{} when clear. We write $a\prec_{\ALG} b$ if $a$ precedes $b$ in the current list maintained by \ALG{}. A request $\sigma_i$ is \emph{pre-served} for an online algorithm \ALG{} at time $t$ if \OPT{} has served $\sigma_i$ by time $t$ but \ALG{} has not. We write $t_i^{\OPT{}}$ and $t_i^{\ALG{}}$ for the times at which \OPT{} and \ALG{} serve $\sigma_i$, respectively, and call the interval $[t_i^{\OPT{}},t_i^{\ALG{}})$ the \emph{pre-service interval} of $\sigma_i$. We write $y_i$ for the position of $e_i$ in \OPT{}'s list at time $t_i^{\OPT{}}$, and $x_i$ for the position of $e_i$ in \ALG{}'s list at time $t_i^{\ALG{}}$.


We analyze our algorithms using a different potential function $\Phi$ for each algorithm. To prove a competitive ratio of $c$, we establish $\Delta\ALG{}+\Delta\Phi\le c\cdot \Delta\OPT{}$ for every event, taking expectations in the randomized case.

\section{Deterministic Algorithms for Time Windows}\label{sec:det}

Our deterministic algorithm builds on the classic move-to-front rule. However, we must also decide when and in what order to serve requests. We first show that several natural choices lead to competitive ratios of $\Omega(n)$. 

\subsection{Limits of simple serving orders}\label{sec:simple-orders}

\begin{restatable}{lemma}{lemSimpleOrders}
\label{lem:simple-orders}
\label{lem:arrival-deadline}
\label{lem:clear-front}
In the time-window model for list update, each of the following deterministic
serving rules has competitive ratio $\Omega(n)$, where $n$ is the list length:
\begin{enumerate}
    \item[(a)] serving every request consistently in either arrival order or
    deadline order;
    \item[(b)] waiting until the next deadline $d_i$, then serving all alive
    requests whose elements precede $e_i$ in the current list, followed by
    $\sigma_i$.
\end{enumerate}
\end{restatable}



See Appendix~\ref{app:simple-orders} for proofs. These examples show that a good algorithm can rely neither on timing alone nor on greedily serving all currently cheap requests; it must balance patience against a selective use of list positions.







\subsection{First-Element Removal and its bounded-overlap analysis}
\label{sec:fer-def}\label{sec:fer-upper}

First-Element Removal (\FER) is given in \cref{alg:fer}. It acts only at deadlines; before serving the expiring request, it clears alive requests for the element currently at the head of the list. We call this the \emph{head-clearance} step. Each such access costs $1$ and leaves the list unchanged. It then serves the expiring request and moves its element to the front.

\begin{algorithm}[t]
\caption{First-Element Removal ($\FER$)}\label{alg:fer}
\begin{algorithmic}[1]
\Statex \textbf{Upon} the deadline of an unserved request $\sigma_i$:
\While{there is an alive unserved request $\sigma_j \ne \sigma_i$ with
$\pos(e_j) = 1$}
  \State serve $\sigma_j$ by accessing $e_j$
\EndWhile
\State serve $\sigma_i$ by accessing $e_i$, then move $e_i$ to the front
of the list
\end{algorithmic}
\end{algorithm}

All actions at a deadline are instantaneous and executed in the stated order, as described in \cref{sec: prelim}. Thus head-clearance does not cause deadline violations. If several deadlines occur at the same time, \FER{} processes the corresponding requests in a fixed order.


We now upper-bound \FER{} in terms of the input's overlap number $k$.

\begin{restatable}{theorem}{thmFERUpper}\label{thm:fer-upper}
For every $k \ge 2$ and every $k$-overlap input $\sigma$,
\[
\FER(\sigma) \;\le\; \Bigl(3 - \frac{2}{k}\Bigr)\,\OPT(\sigma).
\]
\end{restatable}

\subsubsection{Reductions for the analysis of FER}\label{app:reductions}

Throughout the proof of Theorem~\ref{thm:fer-upper} we make the following standing assumptions, without loss of generality:
\begin{itemize}
\item[\textbf{(A1)}] the input is $k$-overlap for some $k \ge 2$ and has no \textit{self-overlap}; 
\item[\textbf{(A2)}] $\OPT$ uses no free exchanges; 
\item[\textbf{(A3)}] $\FER$ serves no requests in its head-clearance steps. As a result, every $\FER$ action is a deadline access followed by a move to the front.
\end{itemize}
The next two lemmas justify (A1)--(A3). Lemma \ref{lem:no-free} justifies (A2) without changing the input; Lemma \ref{lem:reduction} deletes requests to obtain an input satisfying (A1) and (A3) with no decrease in the cost ratio $\FER/\OPT$. Both reductions preserve the $k$-overlap property.

\begin{lemma}\label{lem:no-free}
For any input $\sigma$ there exists an optimal schedule that uses no free exchanges.
\end{lemma}

\begin{proof}
Take an optimal schedule. Suppose it uses free exchange: it serves a request to element $x$ at time $t$ by accessing $x$ at position $p$ and then moves $x$ forward $i$ positions for free. Replace this action by first moving $x$ forward using $i$ paid exchanges, then accessing it at position $p-i$, with both steps performed at time $t$. The replacement costs $i+(p-i)=p$ and leaves the list in the same order, preserving feasibility and all subsequent actions. Repeating this replacement eliminates all free exchanges while preserving optimality.
\end{proof}

For the following reduction, we assume that requests released at time $t$ are available before $\FER$ takes action at $t$. Fix an order for simultaneous deadlines and retain this order when requests are deleted.

\begin{lemma}\label{lem:reduction}
Let $\sigma$ be a nonempty input. Simulate $\FER$ on $\sigma$. Whenever $\FER$ processes an unserved request $\sigma_k$ at its deadline, mark for deletion all other alive requests whose element is either $e_k$ or the current head. Let $\sigma'$ consist of the unmarked requests.

Then $\sigma'$ has no self-overlap, $\FER$ never serves a request through head clearance on $\sigma'$, and
\[
\frac{\FER(\sigma)}{\OPT(\sigma)}
\le
\frac{\FER(\sigma')}{\OPT(\sigma')}.
\]
\end{lemma}

\begin{proof}
Let $m'$ be the number of marked requests for deletion. We first compare $\FER$'s executions on $\sigma$ and $\sigma'$.

Every marked request is served at cost $1$ without changing the list. A request marked because its element is currently at the head is served immediately during head clearance. Otherwise, it is a request to the element $e_k$ of the current request $\sigma_k$. After serving $\sigma_k$, $\FER$ moves $e_k$ to the front. At the next deadline at which $\FER$ acts, the marked request is either served during head clearance or is itself the expiring request. In either case, it is served while $e_k$ is still at the front.

We must also account for deadlines that disappear when marked requests are deleted. If a marked request is itself the expiring request that causes $\FER$ to act, every request served during head clearance at that time is also marked. Thus removing that deadline removes only services of marked requests. 

An induction over the original execution now shows that $\FER$ serves the surviving requests in the same order and from the same list positions on $\sigma'$ as on $\sigma$. Consequently,
$\FER(\sigma)=\FER(\sigma')+m'.$

By Lemma~\ref{lem:no-free}, choose an optimal schedule for $\sigma$ with no free exchanges. Delete its accesses to the marked requests while retaining all paid exchanges. This preserves the list evolution and gives a feasible schedule for $\sigma'$, saving at least one unit per deleted request. Hence
$\OPT(\sigma)\ge\OPT(\sigma')+m'.$

The first deadline trigger survives, so $\sigma'$ is nonempty.
Since $\FER(\sigma')\ge\OPT(\sigma')>0$,
\[
\frac{\FER(\sigma)}{\OPT(\sigma)}
\le
\frac{\FER(\sigma')+m'}{\OPT(\sigma')+m'}
\le
\frac{\FER(\sigma')}{\OPT(\sigma')}.
\]

We finally verify that $\sigma'$ has no self-overlap. Suppose two surviving requests to the same element have intersecting windows. At the start of the earlier of their deadline steps, the other request has arrived and is still unserved. It would therefore be marked, a contradiction. We also check that in $\sigma'$, every head-clearance request in the original execution was marked. Thus $\FER$ serves every surviving request in $\sigma'$ in its own deadline step, with no head-clearance accesses. 
\end{proof}

\subsubsection{Proof of Theorem~\ref{thm:fer-upper}}\label{app:fer-upper}

Recall that $\sigma_i=(e_i,r_i,d_i)$ requests element $e_i$ during the window $[r_i,d_i]$. We work under assumptions (A1)--(A3). Thus $\FER$ serves each request at its deadline, and $\OPT$ changes its list only by paid exchanges. Compare the two schedules in chronological order; at equal timestamps, place $\OPT$'s actions before $\FER$'s, preserving each schedule's internal order. Recall that $t_i^{\OPT}$ and $t_i^{\FER}=d_i$ denote the times at which the two schedules serve $\sigma_i$. A request is \emph{pre-served} after its $\OPT$ access and until its $\FER$ access in this ordering. Let $y_i$ be the position of $e_i$ when $\OPT$ serves $\sigma_i$. 

We first motivate the potential function we use. The difficulty is that $\FER$ may serve a request long after $\OPT$, when the lists have changed. Our aim is to account for these changes when using $\OPT$'s earlier payment to pay for $\FER$'s later service. Recall that we write $a\pF b$ and $a\pO b$ when $a$ precedes $b$ in the current lists of $\FER$ and $\OPT$, respectively. An \emph{inversion} is an unordered pair $\{a,b\}$ with $a \pF b$ and $b \pO a$. We use the number of inversions $\phi$ to compare $\FER$'s list with $\OPT$'s.

Consider $\FER$'s service of $\sigma_i$. Let $x_i$ be the position of $e_i$ immediately before this access, and let $A_i$ be the elements preceding $e_i$ in both lists. Moving $e_i$ to the front creates $|A_i|$ inversions and removes $x_i-1-|A_i|$ inversions. Thus each common predecessor contributes two units to the access cost plus inversion change, while the contributions of other predecessors cancel. Let $F_i$ be the set of $y_i-1$ elements preceding $e_i$ in $\OPT$'s list when it served $\sigma_i$. Then
\[
x_i+\Delta\phi=1+2|A_i|\le 2y_i-1+2|A_i\setminus F_i|.
\] 
This bound separates the cost into a term determined by $\OPT$'s earlier access and two additional units for each element in $A_i\setminus F_i$. Our potential will therefore store credit when $\OPT$ serves a request and collect additional credit as the relative orders change before $\FER$ serves it.

We now formally define the potential function. Set $C=3-2/k$, and let
\[
  \Phi=\phi+
  \sum_{i:\,\sigma_i\text{ pre-served}}(Cy_i+\mu_i+\nu_i).
\]
When $\OPT$ serves $\sigma_i$, create its deposit $Cy_i$ and set $\mu_i=\nu_i=0$. While $\sigma_i$ is pre-served, update the accounts by the rules below. When $\FER$ serves $\sigma_i$, the request ceases to be pre-served, so its summand $Cy_i+\mu_i+\nu_i$ is removed from $\Phi$. The removal subtracts the deposit and account balances immediately before this service, without changing the historical access position $y_i$.

\begin{itemize}
\item If $\OPT$ swaps $e_i$ backward past an adjacent element $w$, increase $\mu_i$ by
\[
  \Delta\mu_i=
  \begin{cases}
    C-1,& \text{if }e_i\pF w,\\
    2,& \text{if }w\pF e_i.
  \end{cases}
\]
\item If $\FER$ moves an element $w$ from behind $e_i$ to the front, and $w\pO e_i$, increase $\nu_i$ by $\Delta \nu_i = 3-C$.
\end{itemize}

The accounts $\mu_i$ and $\nu_i$ start at zero and only increase. Since the deposits $Cy_i$ and the inversion count $\phi$ are also nonnegative, we have $\Phi\ge0$. It therefore suffices to show, for each action,
\begin{equation}\label{eq:perevent}
  \Delta\FER+\Delta\Phi\le C\,\Delta\OPT.
\end{equation}

\begin{lemma}\label{lem:onepre}
Each element has at most one pre-served request. While $\sigma_i$ is pre-served, $\FER$ does not access $e_i$ before serving $\sigma_i$.
\end{lemma}
\begin{proof}
While $\sigma_i$ is pre-served, the current time lies between $\OPT$'s and $\FER$'s service times, both within $[r_i,d_i]$. By (A1), no other request to $e_i$ has a window containing this time. Hence no other request to $e_i$ can be pre-served or served by $\FER$ during this interval.
\end{proof}



\begin{lemma}\label{lem:crossing}
Immediately before $\FER$ serves $\sigma_i$,
\[
  \mu_i+\nu_i\ge2|A_i\setminus F_i|.
\]
\end{lemma}
\begin{proof}
Fix $w\in A_i\setminus F_i$. Hence, $e_i\pO w$ at $t_i^{\OPT}$, and both  $w\pO e_i$ and $w\pF e_i$ hold at $t_i^{\FER}$. By (A2), the order of $w$ and $e_i$ in $\OPT$ can change only when $\OPT$ swaps this pair, so a paid $\OPT$ swap exists moving $e_i$ backward past $w$. Consider the last such $\OPT$ paid swap at time $\tau_w\in [t_i^{\OPT}, t_i^{\FER}]$. Then, observe that $w \pO e_i$ from $\tau_w$ to $t_i^{\FER}$: otherwise another such swap would be needed to restore their final order.

By our definition of the potential function, if $w\pF e_i$ at $\tau_w$, $\mu_i$ receives two units already. Otherwise, it receives $C-1$. In the latter case, $\FER$ must subsequently move $w$ past $e_i$, since $w\pF e_i$ at $t_i^{\FER}$ and $\FER$ does not access $e_i$ in between (Lemma~\ref{lem:onepre}). At this move, $w$ still precedes $e_i$ in $\OPT$ by our choice of $\tau_w$, so $\nu_i$ increases by $3-C$, $3-C+C-1 = 2,$ again giving two units. Different elements $w$ use different pair swaps and different accesses, so no increment is counted twice. Summing over all $w$ proves the claim.
\end{proof}

We finally check that \eqref{eq:perevent} holds under the three events below. Here $\Delta\FER$ and $\Delta\OPT$ are the costs incurred in the action, and $\Delta\Phi$ is the change in potential.

\textbf{When $\OPT$ serves $\sigma_i$}, it pays $y_i$ and changes no list order. Only the deposit is created, so $\Delta\FER+\Delta\Phi=Cy_i=C\,\Delta\OPT$.

\textbf{When $\OPT$ makes a paid adjacent swap}, it pays 1. If the element moved backward has no pre-served request, no account changes and $\Delta\Phi\le1\le C$. Otherwise, by Lemma~\ref{lem:onepre}, exactly one account $\mu_i$ changes. Writing $w$ for the other element, the total change is
\[
  \Delta\Phi=
  \begin{cases}
    1+(C-1)=C,& e_i\pF w,\\
    -1+2=1\le C,& w\pF e_i.
  \end{cases}
\]

Thus \eqref{eq:perevent} holds for every $\OPT$ action.

Finally, \textbf{when $\FER$ serves $\sigma_i$}, its access and move to the front contribute $2|A_i|+1$ to $\Delta\FER+\Delta\phi$, and its own accounts release $Cy_i+\mu_i+\nu_i$. The move may also increase the $\nu$-accounts of other pre-served requests. Their windows and $\sigma_i$'s window all contain the
current time, so there are at most $k-1$ such other requests.
Each account increases by at most $3-C$. Using $|A_i|\le y_i-1+|A_i\setminus F_i|$ and
Lemma~\ref{lem:crossing} gives
\begin{align*}
  \Delta\FER+\Delta\Phi
  &\le 2|A_i|+1-Cy_i-(\mu_i+\nu_i)+(k-1)(3-C)\\
  &\le 2\bigl(y_i-1+|A_i\setminus F_i|\bigr)+1-Cy_i\\
  &\qquad -2|A_i\setminus F_i|+(k-1)(3-C)\\
  &=(2-C)y_i-1+(k-1)(3-C)\\
  &=\left(1-\frac2k\right)(1-y_i)\le0,
\end{align*}
since $y_i\ge1$. Here $\Delta\OPT=0$, completing the proof of
\eqref{eq:perevent}. Summing over all actions and using
$\Phi_{\rm start}=0$ and $\Phi_{\rm end}\ge0$ yields
\[
  \FER(\sigma)\le\left(3-\frac2k\right)\OPT(\sigma).
\]

\begin{remark}[Choice of the constant]\label{rem:forced}
For $2\le C\le3$, the final estimate $(2-C)y_i-1+(k-1)(3-C)$ is maximized at $y_i=1$, where it equals $k(3-C)-2$. It is therefore nonpositive for every $y_i\ge1$ exactly when $C\ge3-2/k$. We choose the smallest such value, $C=3-2/k$.
\end{remark}

\begin{remark}Theorem~\ref{thm:fer-upper} implies that $\FER$ is $3$-competitive for unrestricted inputs. For $k=2$, it is $2$-competitive, matching the deterministic lower bound inherited from classical list update.
\end{remark}


\subsection{Lower bound for FER and local lazy MTF rules}\label{sec:fer-lower}

First we give a lower bound for a natural, broader class of MTF-based algorithms. 



\begin{restatable}[$h$-local-lazy MTF]{definition}{defLazy}\label{def:lazy}
Let $h\in\{0,1,\ldots,n-1\}$. A deterministic online algorithm is
\emph{$h$-local-lazy MTF} if it (i) uses no paid exchanges, (ii) moves every
accessed element to the front, and (iii) serves a request before its deadline
only when its element is in one of the first $h$ positions.
\end{restatable}

The definition permits, but does not require, early service in the first $h$ positions. MTF at deadlines is $h$-local-lazy for every $h$, and \FER{} is $h$-local-lazy for every $h\ge1$.

\begin{restatable}{theorem}{thmLazyLB}\label{thm:lazy-lb}
Fix integers $n\ge3$ and $1\le h<n/2$. Every deterministic
$h$-local-lazy MTF algorithm $A$ admits an input $\sigma$ without self-overlap
such that
\[
\frac{A(\sigma)}{\OPT(\sigma)}
\ge
\frac{n}{\sqrt{2n(n-h)}-n+h+1}.
\]
\end{restatable}

The construction and offline schedule appear in
Appendix~\ref{app:shielded}. When $h=o(n)$, it tends to $1+\sqrt2$. This theorem is a lower bound on worst-case cost ratios. We further show that our 3-competitive analysis is tight. The proof is deferred to Appendix~\ref{app:tw-uniform-local}. 

\begin{theorem}\label{thm:tw-uniform-local}
The competitive ratio of \FER{} is $3$.
\end{theorem}

\section{Randomized Algorithms for Time Windows}\label{sec:rand}


We adapt the classical BIT algorithm to time windows. $\WBIT$ serves every request at its deadline and applies BIT's randomized move rule (Algorithm~\ref{alg:wbit}). It initially chooses independent uniform bits and fixes any ordering of simultaneous deadlines independently of them. Against an oblivious adversary, $\WBIT$ is $7/4$-competitive on all $2$-overlap inputs, matching classical BIT's guarantee~\cite{counterBIT} and beating the deterministic lower bound of $2$. On inputs without self-overlap, our bound improves on $\FER$'s guarantee for $2\le k\le7$ and matches it at $k=8$.





\begin{algorithm}[t]
\caption{Bit Flipping for List Update with Time Windows ($\WBIT$)}\label{alg:wbit}
\begin{algorithmic}[1]
\Statex For each $e \in L$, choose $\bit(e) \in \{0,1\}$ independently
and uniformly at random.
\Statex \textbf{Upon} the deadline of an unserved request $\sigma_i$:
\State serve $\sigma_i$ by accessing $e_i$
\If{$\bit(e_i) = 0$}
  \State move $e_i$ to the front of the list free of charge
\EndIf
\State invert $\bit(e_i)$
\end{algorithmic}
\end{algorithm}




\begin{restatable}{theorem}{thmTWBIT}\label{thm:tw-wbit-improved}
For every integer $k\ge2$ and every input $\sigma$ of overlap at most $k$
with no self-overlap, against an oblivious adversary, $\WBIT$ satisfies
\[
 \E[\WBIT(\sigma)]\le \parenth{\frac{13}{4}-\frac{9}{2(k+1)}}\OPT(\sigma) \ .
\]
 When $k=2$, the no-self-overlap assumption can be omitted.
\end{restatable}

\subsection{Proof of~\cref{thm:tw-wbit-improved}}
Fix $k\ge2$ and an input of overlap at most $k$ with no self-overlap. Fix an optimal schedule independently of the initial bits. 

We first recall previously established results and lemmas. By \cref{lem:no-free}, we may assume that $\OPT$ uses no free exchanges. We use the notation of Section~2.3 with $\mathrm{ALG}=\WBIT$, so since $\WBIT$ always serves a request at its deadline, $t_i^{\WBIT}=d_i$. We continue to use the convention in \cref{app:fer-upper} to order the actions of $\OPT$ and $\WBIT$ when they happen at the same time step.
\cref{lem:onepre} applies here as well: each element has at most one pre-served request, and $\WBIT$ does not access $e_i$ during $\sigma_i$'s pre-service interval before serving $\sigma_i$. Indeed, any other such request would have a window
intersecting $[r_i,d_i]$.

The event sequence and the pre-service intervals are independent of the initial bits. At any fixed event, the current bit of an element is its initial bit XOR the parity of a fixed number of preceding accesses. The current bits are therefore independent and uniform. They need not be independent of $\WBIT$'s list order.

\subsubsection{The potential function.} As in \cref{sec:fer-upper}, the potential function deposits credit when $\OPT$ pre-serves a request and accumulates credits in the $\mu$ and $\nu$ accounts to pay for $\WBIT$'s later service of the request. 

Set
\[
 C=\frac{13}{4}-\frac{9}{2(k+1)},
 \qquad \delta=\frac{9}{2(k+1)}.
\]
We use the potential function
\[
 \Phi=\varphi+\sum_{i:\,\sigma_i\text{ pre-served}}
 \left(Cy_i+\sum_{f\ne e_i}(\mu_{i,f}+\nu_{i,f})\right).
\]
Here $\varphi$ is the weighted inversion term, $Cy_i$ is the deposit for $\sigma_i$, and $\mu_{i,f},\nu_{i,f}$ are its credits associated
with element $f$. We motivate and define these terms in more detail next.

Recall that an inversion is an unordered pair of distinct elements $\{e,f\}$ where $e\pO f$ and $f\pW e$. We now give every inversion pair a weight defined with $\operatorname{bit}(e)$. Each inversion $(e, f)$ has weight $1+b_e,$ where $b_e = \operatorname{bit}(e)$, and we define $$\varphi = \sum_{\substack{e\prec_{\OPT}f\\f\prec_{\WBIT}e}}
 (1+b_e).$$
The weights attached to the inversions allow a $\WBIT$'s access to release potential even when $\operatorname{bit}(e)=1$. In this case where the element requested is not moved to the front and the inversion remains, $\operatorname{bit}(e)$ flips from 1 to 0, still releasing one unit to pay for the access.

The deposit $Cy_i$ carries a multiple of $\OPT$'s payment for $\sigma_i$ forward to $\WBIT$'s later service. As in~\cref{app:fer-upper}, let $F_i$ be the $y_i-1$ predecessors of $e_i$ when $\OPT$ serves $\sigma_i$. Elements outside $F_i$ may move ahead of $e_i$ in $\OPT$ before $\WBIT$ serves it. The accounts $\mu_{i,f}$ and $\nu_{i,f}$ collect credits to pay for these new predecessors for each request--element pair $(\sigma_i,f)$, with $f\ne e_i$: $\mu_{i,f}$ records credits from $\OPT$'s backward swaps of $e_i$ across $f$, and $\nu_{i,f}$ records credits from $\WBIT$'s forward moves of $f$ across $e_i$. We keep the credits for each $f$ separate because its contribution to the later access calculation depends on its own bit value.

When $\OPT$ serves $\sigma_i$, create a deposit $Cy_i$ and initialize $\mu_{i,f}=\nu_{i,f}=0$ for every $f\ne e_i$. During $e_i$'s pre-service interval, update the accounts as follows:
\begin{itemize}
\item When $\OPT$ moves $e_i$ backward across $f$ in a paid adjacent swap, add to $\mu_{i,f}$ the amount
\[
 \Delta \mu_{i,f}=
 \begin{cases}
 C+1,&f\prec_{\WBIT} e_i,\\
 C-1-b_f,&e_i\prec_{\WBIT} f.
 \end{cases}
\]
\item When a $\WBIT$ access to $f$ moves $f$ forward across $e_i$,
add $\delta$ to $\nu_{i,f}$ if $f\prec_{\OPT} e_i$.
\end{itemize}
When $\WBIT$ serves $\sigma_i$, remove its deposit $Cy_i$ and all credits $\mu_{i,f},\nu_{i,f}$ for $f\ne e_i$ from the potential.

\subsubsection{Event analysis.}
We show at every fixed event that
\[
 \E[\Delta\WBIT+\Delta\Phi]\le C\,\Delta\OPT.
\]
Here each $\Delta$ denotes the change during the event under consideration, and bits and list orders are evaluated immediately before that event. All expectations are over the initial bits, without conditioning on $\WBIT$'s current list or its preceding random choices.

\textbf{When $\OPT$ serves $\sigma_i$}, it pays $y_i$ and changes
no list order. Only the deposit
$Cy_i$ is created. Thus
$\Delta\WBIT+\Delta\Phi=Cy_i=C\Delta\OPT$.

\textbf{When $\OPT$ makes a paid adjacent swap}, there are two cases depending on whether the backward-moving element has a pre-served request. Suppose first that $\OPT$ moves an element $e_i$ with pre-served request backward across $f$. No self-overlap ensures that only one request to $e_i$ is pre-served, so only $\mu_{i,f}$ receives an increment of
$\Delta\mu_{i,f}$. If $e_i\prec_{\WBIT}f$, this creates a new inversion with weight $1+b_f$, and
\[
 \Delta\Phi=\Delta\varphi+\Delta\mu_{i,f}
 =(1+b_f)+(C-1-b_f)=C.
\]
Otherwise, if $f\prec_{\WBIT}e_i$, the swap removes an inversion of weight
$1+b_{e_i}$, and
\[
 \Delta\Phi=\Delta\varphi+\Delta\mu_{i,f}
 =-(1+b_{e_i})+(C+1)=C-b_{e_i}\le C.
\]
If the backward-moving element has no pre-served request, the only change in the potential function is the inversion. The change is at most $1+b_f$, with $f$ the forward-moving element, so its expectation is at most $3/2\le C$. Since $\Delta\WBIT=0$ and $\Delta\OPT=1$, the
required inequality holds in every case.

\textbf{When $\WBIT$ serves $\sigma_i$}, it incurs the access cost and updates the inversion term, while also removing the deposit
$Cy_i$ and all credits $\mu_{i,f},\nu_{i,f}$ for $f\ne e_i$. 

\begin{align*}
 \Delta\WBIT+\Delta\Phi
 =\Delta\WBIT+\Delta\varphi-Cy_i
   -\sum_{f\ne e_i}(\mu_{i,f}+\nu_{i,f})\quad+
   \sum_{\substack{j\ne i\\\sigma_j\text{ pre-served}}}
   \Delta\nu_{j,e_i}.
\end{align*}
Here $\mu_{i,f}$ and $\nu_{i,f}$ are the balances immediately before the service. The last sum records the increase in $\nu_{j, e_i}$ when $\WBIT$ moves $e_i$ forward pass other elements $e_j$ that has pre-served requests.

We first determine the access cost together with $\Delta \varphi$.
For distinct elements $e,f$, define
\[
 X_e(f)=\ind{f\prec_{\WBIT} e}
       \bigl[1+(1-b_e)(1+b_f)\bigr].
\]

\begin{lemma}\label{lem:wbit-identity}
During a $\WBIT$ access to $e$, with $\OPT$'s list held fixed,
\begin{equation}\label{eq:tw-wbit-identity}
 \Delta\WBIT+\Delta\varphi
 =1+\sum_{f\prec_{\OPT} e}X_e(f).
\end{equation}
At every fixed event and for every fixed pair $e\ne f$,
\begin{equation}\label{eq:wbit-pair-mean}
 \E X_e(f)\le\frac74=:\alpha.
\end{equation}
\end{lemma}
\begin{proof}
All bits below are evaluated immediately before the access.
Each element $f$ preceding $e$ in $\WBIT$ contributes one unit
to the access cost.

If $e\prec_{\OPT}f$, this unit is canceled by a one-unit decrease
in $\varphi$: the access removes a weight-one inversion when
$b_e=0$, or reduces its weight from two to one when $b_e=1$.

If instead $f\prec_{\OPT}e$, the pair is initially not an inversion.
When $b_e=0$, moving $e$ to the front creates an inversion of
weight $1+b_f$, since $f$ is earlier in $\OPT$.
When $b_e=1$, no inversion is created. Including the access-cost
unit, the contribution is therefore
\[
 1+(1-b_e)(1+b_f)=X_e(f).
\]

An element $f$ following $e$ in $\WBIT$ contributes no access cost,
and its relative order with $e$ does not change. Any inversion
between them has weight $1+b_f$, which is also unchanged.
Pairs not containing $e$ are unaffected. Summing these contributions
and adding the unit access cost for $e$ itself proves the identity.

For the expected bound, the indicator in $X_e(f)$ is at most one
and its multiplier is nonnegative. At a fixed event, the bits
of any fixed pair are independent and uniform, so
\[
 \E[X_e(f)]
 \le 1+\E[(1-b_e)(1+b_f)]
 =1+\frac12\cdot\frac32
 =\frac74.
\]
This does not require the bits to be independent of $\WBIT$'s
list order.
\end{proof}

Observe that 
 $$\sum_{f\prec_{\OPT} e_i}X_{e_i}(f) = \sum_{f\prec_{\OPT} e_i, f\in F_i}X_{e_i}(f)+ \sum_{f\prec_{\OPT} e_i, f\notin F_i}X_{e_i}(f). $$  By Lemma~\ref{lem:wbit-identity}, the first term is at most $\alpha(y_i-1)$ in expectation at $\WBIT$'s service of $\sigma_i$. We now tackle the second term. Let $N_i$ be the elements outside $F_i$ that precede $e_i$ in $\OPT$ when $\WBIT$ serves $\sigma_i$. The following lemma shows that the credits $\mu_{i,f}+\nu_{i,f}$
 cover the contribution of each new predecessor $f\in N_i$ in expectation.

\begin{lemma}\label{lem:wbit-coverage}
For every $f\in N_i$, immediately before $\WBIT$ serves $\sigma_i$,
\[
 \E X_{e_i}(f)\le\E[\mu_{i,f}+\nu_{i,f}].
\]
\end{lemma}

\begin{proof}
Fix $f\in N_i$. Hence,
$e_i\prec_{\OPT}f$ when $\OPT$ serves $\sigma_i$, whereas
$f\prec_{\OPT}e_i$ immediately before $\WBIT$ serves $\sigma_i$.
The relative order of $e_i$ and $f$ in $\OPT$ can change only
through a paid swap of this pair, so a paid swap moving $e_i$
backward past $f$ exists during the pre-service interval.
Consider the last such swap, and denote this event by $\tau_f$.
Then $f\prec_{\OPT}e_i$ from immediately after $\tau_f$ until
$\WBIT$ serves $\sigma_i$: otherwise another such swap would
be needed to restore their final order.

In this proof, $\Delta\mu_{i,f}$ denotes the increment at $\tau_f$,
and $b_{e_i},b_f$ denote the bits at $\tau_f$.
The quantities $X_{e_i}(f)$, $\mu_{i,f}$, and $\nu_{i,f}$ are
evaluated immediately before $\WBIT$ serves $\sigma_i$.
By no self-overlap, $\WBIT$ does not access $e_i$ in between,
so its bit does not change. Every later $\WBIT$ move of $f$ past $e_i$
adds $\delta$ to $\nu_{i,f}$, since $f$ remains ahead in $\OPT$.

The event $\tau_f$ and the number of later accesses to $f$ are
independent of the initial bits. Each earlier increment to $\mu_{i,f}$ is at least $C-1$ minus
the bit of $f$ at that swap. Since this bit is uniform, the
increment has expectation at least $C-3/2\ge0$.
The set of earlier swaps depends only on $\OPT$ and the input,
so their increments have nonnegative total expectation.
It therefore suffices to prove
\begin{equation}\label{eq:tw-wbit-cover}
 \E[X_{e_i}(f)]
 \le \E[\Delta\mu_{i,f}+\nu_{i,f}].
\end{equation}
We consider $\WBIT$'s accesses to $f$ after $\tau_f$ and before
$\WBIT$ serves $\sigma_i$.

\emph{No later access to $f$.}
Neither bit nor the relative order changes.
If $f$ precedes $e_i$ in $\WBIT$, then
$\Delta\mu_{i,f}=C+1$ and $X_{e_i}(f)\le 2+b_f$.
Otherwise, $\Delta\mu_{i,f}=C-1-b_f$ and $X_{e_i}(f)=0$.
No further credit is added to $\nu_{i,f}$ before $\WBIT$ serves
$\sigma_i$, but any earlier credits remain. Since these credits
are nonnegative, both cases give
\[
 \E[\Delta\mu_{i,f}+\nu_{i,f}-X_{e_i}(f)]
 \ge C-1-\E[b_f]
 =C-\frac32\ge0.
\]

\emph{One later access to $f$.} 
If $b_f=0$, this access moves $f$ to the front of $\WBIT$ and flips its bit
to one, so $X_{e_i}(f)=3-2b_{e_i}$.
If $f\pW e_i$ at $\tau_f$, the swap contributed
$C+1$. Otherwise, it contributed $C-1$, and $f$ is moved forward across $e_i$, adding $\delta$ to $\nu_{i,f}$.
Since $\delta\le3/2$, in either case
\[
 \Delta\mu_{i,f}+\nu_{i,f}-X_{e_i}(f)
 \ge \min\{C+1, C-1+\delta\} - (3-2b_{e_i}) =C+\delta-4+2b_{e_i}.
\]

If $b_f=1$, the access leaves $f$ in place and flips its bit to zero.
If $e_i\pW f$, then $\Delta\mu_{i,f}=C-2$ and
$X_{e_i}(f)=0$. If $f\pW e_i$, then
$\Delta\mu_{i,f}=C+1$ and $X_{e_i}(f)=2-b_{e_i}\le2$.
Thus in either case, $\Delta\mu_{i,f}+\nu_{i,f}-X_{e_i}(f)\ge C-2$.

These bounds hold for either relative order at $\tau_f$.
The bits there are independent and uniform, so
\[
 \E[\Delta\mu_{i,f}+\nu_{i,f}-X_{e_i}(f)]
 \ge \frac12(C+\delta-4+1)+\frac12(C-2)
 =C-\frac52+\frac{\delta}{2}\ge0.
\]

\emph{At least two later accesses to $f$.} This guarantees that $f$ is moved to the front during one of the first two accesses, regardless of its bits at $\tau_f.$ If $f$ was behind $e_i$ at $\tau_f$, this move to front crosses $e_i$ and adds $\delta$ to $\nu_{i,f}$, giving
\[
 \Delta\mu_{i,f}+\nu_{i,f}\ge C-1-b_f+\delta.
\]
If $f$ was already ahead, the swap contributed $C+1$, which also satisfies this bound because $\delta\le3/2$. Consequently,
\[
 \E[\Delta\mu_{i,f}+\nu_{i,f}]
 \ge C-\frac32+\delta
 =\frac74
 \ge \E[X_{e_i}(f)],
\]
where the last inequality follows from Lemma~\ref{lem:wbit-identity}.
\end{proof}

By Lemma~\ref{lem:wbit-coverage}, for each $f\in N_i$, removing $\sigma_i$'s accounts $\mu_{i,f}$ and $\nu_{i,f}$ at the time of $\WBIT$'s service offsets the contribution $X_{e_i}(f)$ in expectation. For each remaining $f\ne e_i$, the account $\mu_{i,f}$ is a sum of increments from $\OPT$ swaps during $\sigma_i$'s pre-service
interval. These swaps are fixed independently of the bits, and each increment has expectation at least $C-3/2\ge0$. Also, $\nu_{i,f}\ge0$. Since $N_i$ is fixed independently of the bits, removing all these remaining accounts cannot increase the
expected charge. Including the result from~\cref{lem:wbit-identity} and subtracting the deposit $Cy_i$ therefore gives
\[
 \E\left[
 \Delta\WBIT+\Delta\varphi-Cy_i
 -\sum_{f\ne e_i}(\mu_{i,f}+\nu_{i,f})
 \right]
 \le 1+\alpha(y_i-1)-Cy_i.
\]
It remains to account for the credits added to other pre-served requests during this access. We show that their expected total is at most $\delta(k-1)/2=C-1$. If $\WBIT$ moves $e_i$ forward across $e_j$ and $e_i\prec_{\OPT}e_j$, it increases $\nu_{j,e_i}$ by $\delta$. From the $k$-overlap restriction, there are at most $k-1$ other pre-served requests, and each receives at most one increment of $\delta$. These increments occur only if $\WBIT$ moves $e_i$ forward, i.e. if $b_{e_i}=0$, so
\[
 \sum_{\substack{j\ne i:\,\sigma_j\text{ pre-served}}}
 \Delta\nu_{j,e_i}
 \le \delta(k-1)\ind{b_{e_i}=0}.
\]
Since $b_{e_i}$ is uniform at this fixed event, the expected total
increment is at most $\delta(k-1)/2=C-1$.  Combining these increments with the removed deposit
and credits gives
\begin{align*}
 \E[\Delta\WBIT+\Delta\Phi]
 &\le\alpha(y_i-1)+1+\frac{\delta(k-1)}2-Cy_i\\
 &=\alpha(y_i-1)+C-Cy_i\\
 &=(\alpha-C)(y_i-1)\le0.
\end{align*}
Here $\Delta\OPT=0$, so all three event inequalities hold.

Initially the lists agree and all accounts are absent, so $\Phi=0$.
At the end every request has been served by both schedules; all
request accounts have been removed and $\Phi=\varphi\ge0$.
Summing the expected event inequalities gives
\[
 \E[\WBIT(\sigma)+\Phi_{\rm end}]
 \le C\OPT(\sigma),
\]
and hence $\E[\WBIT(\sigma)]\le C\OPT(\sigma)$. This proves the
assertion of Theorem~\ref{thm:tw-wbit-improved} under no self-overlap.

\subsubsection{Choice of the constant.}
The case of at least two later accesses needs $C+\delta\ge13/4$,
while issuing new crossing credits needs
$C\ge1+\delta(k-1)/2$. Equating these gives
\[
 \delta=\frac{9}{2(k+1)},
 \qquad C=\frac{13}{4}-\frac{9}{2(k+1)}.
\]

\subsection{Removing self-overlap when $k=2$}\label{app:removek=2}
For $2$-overlap inputs, the intersection of two windows for the same element cannot meet any third request's window. This allows us to remove self-overlap without changing \WBIT{}'s access sequence or increasing the cost of \OPT{}.

\begin{lemma}\label{lem:wbit-uncross}
For every $2$-overlap input $\sigma$, there is a $2$-overlap input $\sigma'$ with no self-overlap such that $\WBIT(\sigma')=\WBIT(\sigma).$ For every initial bit assignment, $\OPT(\sigma')\le\OPT(\sigma).$
The transformation is independent of the random bits and preserves the initial list.
\end{lemma}

\begin{proof}
    We first construct $\sigma'$. Consider any two overlapping requests $\sigma_i$ and $\sigma_j$ to the same element $e$, with $d_i<d_j$ (assuming endpoints are disjoint, we handle tied endpoints at the end of the proof). Let $a=\min\{r_i, r_j\},$ $b = \max\{r_i, r_j\}$. Choose $c>d_i$ before the first release time or deadline of any request occurring after $d_i$. In particular, $c<d_j$. Replace the windows of $\sigma_i$ and $\sigma_j$ respectively with $[a, d_i]$ and $[c, d_j].$ 
    Apply this replacement to the remaining overlapping pairs in deadline order, and call the resulting input $\sigma'$. Each step preserves $2$-overlap and keeps previously processed requests free of self-overlap, so $\sigma'$ has no self-overlap.

    We next show that a single replacement preserves $\WBIT$'s cost for every initial bit assignment and does not increase $\OPT$'s cost. Applying these comparisons successively proves the lemma.

    Since we preserve all requests' deadlines and their requested elements, we preserve $\WBIT$'s cost: $\WBIT(\sigma) = \WBIT(\sigma').$ 

    Now to compare offline costs, take an optimal schedule for the original input and consider its two accesses serving $\sigma_i$ and $\sigma_j$. Relabel the earlier access
    as the service of $\sigma_i$, and the later access as the service of $\sigma_j$. Since both requests are to the same element $e$, this changes only the request labels, not the actions or their costs.

    For the earlier access, it must lie in $[a, d_i]$: neither access can precede the earliest release time $a$, and at least one must occur by $d_i$ to serve the original $\sigma_i.$ Similarly, the later access must lie in $[b, d_j]$: at least one access must be made at or after $b$, and both requests must be served by $d_j$. If the later access also lies in $[c, d_j],$ then the schedule can also apply to our new input and we're done. Otherwise, postpone this later access, any immediate free move, and all subsequent paid swaps before $c$ to time $c$, preserving their order. By $2$-overlap and the choice of $c$, no third request's window meets $[b,c]$. The postponement therefore delays no other request's service, and $\sigma_j$ is served feasibly at its new release $c$. Since the order of all accesses and moves is unchanged, their costs are unchanged.
    
    The constructed schedule is feasible for the modified input and has cost $\OPT(\sigma)$. Hence the minimum feasible cost for that input satisfies $\OPT(\sigma')\le\OPT(\sigma)$.

    To handle tied endpoints, choose $\varepsilon>0$ small enough that shifting endpoints by less than $\varepsilon$ preserves the order of endpoints that originally occurred at different times. Move the releases at each time $t$ to distinct times in $(t-\varepsilon,t)$, and the deadlines at $t$ to distinct times in $(t,t+\varepsilon)$, ordered according to $\WBIT$'s fixed, bit-independent tie order.
    
    Keep every action of the original $\OPT$ schedule at its original time. Since releases move earlier and deadlines move later, every service remains within its request's window. Thus the perturbation does not increase the optimal cost.
    
    For $\WBIT$, the new deadlines preserve the order of accesses. For any fixed initial bit assignment, the same access sequence produces the same bit flips and list movements. Each access therefore occurs at the same list position and incurs the same cost; only its time changes.
    
    These changes do not increase overlap. Every window present within $(t-\varepsilon,t+\varepsilon)$ already contained $t$ in the original input. Outside these intervals, the windows are unchanged. All endpoints are now distinct, so we can apply the construction above.

\end{proof}

As a result, \WBIT{} is $7/4$-competitive on all $2$-overlap inputs against an oblivious adversary, matching classical BIT's guarantee~\cite{counterBIT} and below the deterministic lower bound of $2$.


\section{The Delay Model}\label{sec: delay}

Throughout this section, delay functions are nondecreasing, eventually divergent, and zero at release. Time in the delay model is continuous. An algorithm may access or swap elements at any nonnegative real time. These actions are instantaneous: they incur cost but take no time. In the delay-unaware setting, an algorithm continuously observes each pending request's current delay but does not know its future delay. In the delay-aware setting, the entire delay function for a request is revealed when the request arrives.

\subsection{A lower bound for delay-unaware algorithms}
\label{delay-lb}


\begin{restatable}{theorem}{thmdelaylb}\label{thm:nba-delay-prefix-lb}
No deterministic delay-unaware algorithm for NBA list update with delays has a competitive ratio smaller than $2+\sqrt{3}$. 
\end{restatable}

\begin{proof}
Fix a deterministic delay-unaware algorithm $A$ and a list with $n$ elements, with $n\ge 2$. We construct inputs with arbitrarily large $m$ requests, so that we rule out an additive constant depending on $n$.

\paragraph{Construction.} We construct requests $\sigma_1,\ldots,\sigma_m$ recursively as $A$ runs. Request $\sigma_i$ asks for element $e_i$, is released at time $r_i$, and has delay function $f_i$. Let $s_i$ be the time at which $A$ serves it. Set $r_1=0$ and $r_i=s_{i-1}+1$ for $2\le i\le m$, and choose $e_i$ to be the last element in $A$'s list immediately before $\sigma_i$ is released. Thus $A$ has at most one pending request at a time. The times $r_i$ and $s_i$ need not be integers; the added unit only separates one request's service from the next release.

The delay function of $\sigma_i$ grows at unit rate until $A$
serves it, remains constant through time $s_m+1$,
and then resumes unit-rate growth:
\[
 f_i(t)=
 \begin{cases}
  t-r_i, & r_i\le t<s_i,\\
  s_i-r_i, & s_i\le t\le s_m+1,\\
  s_i-r_i+t-(s_m+1), & t>s_m+1.
 \end{cases}
\]
Each function is continuous, nondecreasing, zero at release,
and eventually divergent.

Once the construction is complete, the displayed functions define a fixed input for both $A$ and $\OPT$. Because $A$ is deterministic and delay-unaware, specifying the functions' future values does not change the observations or decisions made during the construction. 

We may assume that $A$ serves every released request. Otherwise, suppose it stops at request $\sigma_i$. Release no further requests and let its delay exceed $Cni+b_A(n)$ for any claimed guarantee $A(\sigma)\le C\OPT(\sigma)+b_A(n)$. At that point, let all earlier frozen functions resume unit-rate growth, so every delay function is eventually divergent. An offline schedule can serve the $i$ released requests immediately for total cost at most $ni$. Thus $A$ violates the guarantee if it eventually serves $\sigma_i$, and is infeasible if it never does. Hence every $s_i$ may be assumed finite.

\paragraph{Online cost.} By construction, every requested element is at position $n$ when its request arrives. Until that request is served by $A$, it is the only request available for service. Neither paid swaps before the access or free forward movement after the access could reduce the access cost\footnote{Paid swaps may lower the eventual access position, but each position gained costs one paid swap. Free forward movement occurs only after the access and therefore cannot reduce the cost when serving the current request.}. Thus the access and preceding movement cost at least $n$ per request. These pending intervals are disjoint, so no swap is charged twice. If $\Delta$ denotes the total delay paid by $A$, then
\begin{equation}\label{eq:delay-prefix-online}
 A(\sigma)\ge nm+\Delta.
\end{equation}

\paragraph{Offline cost.} 
We now construct an offline schedule for the input $\sigma$ defined above by $A$. It serves requests to some elements immediately and postpones requests to the others. Postponement lets it serve all requests to one element consecutively at position $1$, while the constant segments of the delay functions prevent any additional delay cost after $A$'s service.

Consider any ordering of the $n$ elements; we will use averaging later to show that some ordering gives the desired cost bound. At time $0$, arrange the offline list in this order using at most $\binom n2$ paid adjacent swaps. The first $\ell$ elements in this order are the ones whose requests we serve immediately at release, $1\le\ell<n$. We will choose $\ell$'s value after calculating the cost. The schedule then proceeds as follows.

\begin{enumerate}
 \item \emph{Before time $s_m+1$,} serve each request to one of the first $\ell$ elements at its release time, without changing the list. Postpone all other requests until time $s_m+1$.

 For a particular ordering, its total cost is $$\sum_{j=1}^\ell j\cdot \text{(\# of requests to the element at position $j$)}.$$

 \item \emph{At time $s_m+1$,} serve all requests postponed previously one at a time. Choose an element with pending requests, move it to the front using at most $n-1$ paid adjacent swaps, and then serve every pending request to it by a separate access at position $1$. Continue with the next element until all postponed requests have been served. All these actions occur at time $s_m+1$, in the stated order, without advancing time. 
 
 Each element is moved at most once, so the total movement cost is at most $(n-\ell)(n-1)$. Each postponed request $\sigma_i$ pays one unit for access and delay $f_i(s_m+1)=s_i-r_i$, equal to its delay under $A$. Its access and delay costs therefore total $1+s_i-r_i$.
\end{enumerate}

The schedule above is feasible for every ordering, but its cost depends on which elements occupy the first $\ell$ positions. To upper-bound $\OPT$, we only need one ordering for which this cost is small. We show that such an ordering exists by averaging the total cost over all orderings with equal weight: at least one ordering costs no more than this average.

Before $s_m+1$, for each request, if its requested element has location $j\leq\ell$, it costs $j$, otherwise it costs 0. For each requested element, it occupies position $j$ in fraction $\frac{1}{n}$ of all possible orderings. Summing over all $m$ requests, the average cost before time $s_m+1$ is
\[
 \frac{m}{n}\sum_{j=1}^{\ell}j
 =\frac{\ell(\ell+1)}{2n}\,m.
\]

In a fraction $1-\ell/n$ of the orderings, the service of request $\sigma_i$ is postponed to $s_m+1$, where its access and delay cost is $1+s_i-r_i$.
Since $\sum_{i=1}^m(s_i-r_i)=\Delta$ (recall $\Delta$ is the total delay paid by $A$), the average access and delay cost at time $s_m+1$ is
\[
 \left(1-\frac{\ell}{n}\right)
 \sum_{i=1}^m(1+s_i-r_i)
 =\left(1-\frac{\ell}{n}\right)(m+\Delta).
\]

The paid swaps, both initially to form the ordering and at time $s_m+1$, cost at most
 $$\binom n2+(n-\ell)(n-1)\le2n^2.$$
This movement bound holds for every ordering.

Adding the three costs bounds the average total cost. Hence
\begin{equation}\label{eq:delay-prefix-offline}
 \OPT(\sigma)\le
 \underbrace{\frac{\ell(\ell+1)}{2n}\,m}_{\text{Step 1}}
 +\underbrace{\left(1-\frac{\ell}{n}\right)(m+\Delta)}_{\text{Step 2: accesses and delay}}
 +\underbrace{2n^2}_{\text{paid swaps}}.
\end{equation}

Now choose $\ell=\lfloor(\sqrt3-1)n\rfloor$. Then
\cref{eq:delay-prefix-offline} gives
\begin{equation}\label{eq:delay-prefix-direct}
 \OPT(\sigma)\le
 \left(2-\sqrt3+\frac1n\right)(nm+\Delta)+2n^2.
\end{equation}
To verify~\eqref{eq:delay-prefix-direct}, bound the coefficients of
$m$ (access costs) and $\Delta$ (delay costs) in
\eqref{eq:delay-prefix-offline}:
\begin{align*}
 \frac{\ell(\ell+1)}{2n}+1-\frac{\ell}{n}
 &=1+\frac{\ell(\ell-1)}{2n}
 \le (2-\sqrt3)n+1,\\
 1-\frac{\ell}{n}&\le2-\sqrt3+\frac1n.
\end{align*}

We next conclude that the competitive ratio has the lower bound of $2+\sqrt{3}.$ Suppose for contradiction that $A$ had competitive factor $C<2+\sqrt3$. This would mean
\[
 A(\sigma)\le C\OPT(\sigma)+b_A(n)
 \qquad\text{for every input }\sigma,
\]
where $b_A(n)$ is finite and may depend on $A$ and $n$, but not on
the request sequence or its length $m$.
Since $C(2-\sqrt3)<1$, choose $n$ large enough that
$C(2-\sqrt3+1/n)<1$, and then keep this $n$ fixed.
Combining the claimed guarantee with~\eqref{eq:delay-prefix-online}
and~\eqref{eq:delay-prefix-direct} would require
\[
 \left[1-C\left(2-\sqrt3+\frac1n\right)\right](nm+\Delta)
 \le 2Cn^2+b_A(n).
\]
Fixing $n$ keeps the entire right-hand side fixed, including the
additive constant. The bracket is positive and $\Delta\ge0$, so
increasing $m$ makes the left-hand side arbitrarily large, a
contradiction. 
\end{proof}

\subsection{Delayed Move-to-Front}\label{delay_6_comp}

We again use move-to-front, but now the accumulated delay determines when to access an element. 
Let $R_e(t)$ denote the set of pending requests to $e$, and let $c_e(t)=\sum_{\sigma_i\in R_e(t)}f_i(t)$ be the total delay across all requests in the set.
\ALGDELAY{}, presented in \cref{alg:delayMTF}, is delay-unaware and serves one request to $e$ as soon as $c_e(t)$ reaches $2/3$ of its current position. It moves $e$ to the front for free and serves all remaining requests on it with a cost of 1 each. 

The accumulated delay cost $c_e(t)$ may reach the triggering condition simultaneously for different elements. We use any fixed tie-breaking, and each time after the move-to-front, the element positions are recomputed and the triggering condition might change accordingly. 

\begin{algorithm}[t]
\caption{Delayed Move-to-Front ($\ALGDELAY$)}\label{alg:delayMTF}
\begin{algorithmic}[1]
\Statex \textbf{Upon} $c_e(t)=\frac23\pos(e)$ for an element $e$:
\State serve one pending request to $e$ by accessing $e$
\State move $e$ to the front of the list free of charge
\State serve each remaining pending request to $e$ by a separate access
\end{algorithmic}
\end{algorithm}

In the delay-aware extension, we also allow left-continuous delay functions with jumps. Since the functions are known at release, \ALGDELAY{} serves at the threshold-crossing time; if a jump would carry the delay above the threshold, it serves at the jump time, before the higher value applies. 

The proof of Lemma~\ref{lem:no-free} also applies to the delay model: the transformation preserves every service time, so all delay costs remain unchanged. Thus, we may assume that \OPT{} uses no free exchanges.


\begin{theorem}\label{thm: delayMTF-appendix}
For continuous, nondecreasing, divergent delay functions with $f_i(r_i)=0$, the $2/3$-threshold algorithm \ALGDELAY{} is $5$-competitive in the delay-unaware model. Its delay-aware extension is also $5$-competitive for left-continuous functions.
\end{theorem}

\begin{proof}
We compare DMF and $\OPT$ on the same input. The two schedules may serve requests at different times and maintain different list orders. We use a potential to account for these differences and prove that DMF's total cost is at most five times $\OPT$'s. We first prove the bound for continuous delay functions in the delay-unaware regime. We then extend the proof to left-continuous delay functions with jumps in the delay-aware regime, where each function is known in full at release.

Before defining the potential, we calculate DMF's cost for each group of requests it serves together. We then construct a comparison schedule $\OPT^*$ with one offline access per group and identify the inequality that the potential must establish.

We designed DMF such that it waits until the total delay of the pending requests to some element $e$ reaches two-thirds of $e$'s current position. It then serves all pending requests to $e$ at that time. We call the set of these actions on $e$ one \emph{service event}. Index the events by $j=1,\ldots,J$. At event $j$, let $e_j$ be the element served, $R_j$ its set of pending requests, $t_j$ the service time, and $x_j$ its position just before service. DMF serves one request by accessing $e_j$ at cost $x_j$, moves $e_j$ to the front for free, and serves each of the other $|R_j|-1$ requests by a separate access at cost $1$. 
We observe that
\begin{equation}\label{eq:delay-trigger-properties}
 \sum_{\sigma_i\in R_j}f_i(t_j)=\frac23x_j,
 \qquad
 \sum_{\sigma_i\in R_j}f_i(u)\ge\frac23x_j
 \quad\text{for every }u>t_j.
\end{equation}
The first equality follows from the definition of the service time, and the second inequality follows from monotonicity.

Thus DMF's cost at event $j$ is at most $\frac53x_j+(|R_j|-1)$. Let $M=\sum_j(|R_j|-1)$ count all the accesses made after moving an element to the front. Summing the event costs gives
\begin{equation}\label{eq:delay-online-cost}
 \ALGDELAY(\sigma)\le\sum_j\frac53x_j+M.
\end{equation}

We now construct $\OPT^*$. By~\cref{lem:no-free}, we may assume that $\OPT$ uses no free exchanges. We can therefore delete an offline access without changing the list or the positions of later accesses. For each set of requests $R_j$, keep only the first access by $\OPT$ that serves a request $\hat\sigma_j$ in $R_j$ and delete all other services to requests in $R_j$. Let $\tau_j$ be the time at which \OPT{} serves $\hat\sigma_j$, and let $y_j$ be the position of $e_j$ in \OPT{} at time $\tau_j$. Keep every paid swap, and preserve the times and order of all retained actions. Call this reduced schedule $\OPT^*$.

In $\OPT^*$, assign the retained access the delay charge
\[
 \begin{cases}
  0, & \tau_j\le t_j,\\
  \frac23x_j, & \tau_j>t_j.
 \end{cases}
\]
$\OPT^*$'s delay defined above never exceeds $\OPT$'s total delay for $R_j$. The zero case is immediate. In the other case, every request in $R_j$ is served by $\OPT$ at or after $\tau_j>t_j$, so their total delay is at least $\sum_{\sigma_i\in R_j}f_i(\tau_j)\ge\frac23x_j$ by \eqref{eq:delay-trigger-properties}.

Consider the cost of $\OPT^*$, also denoted $\OPT^*$. Note that we use this quantity only for comparison; we do not run DMF on a modified input. Exactly $M$ offline accesses were deleted, each costing at least $1$, and neither the remaining access costs nor the delay charges increased. Consequently,
\begin{equation}\label{eq:delay-offline-cost}
 \OPT^*\le\OPT(\sigma)-M.
\end{equation}
Hence, to show that DMF$(\sigma)$ is 5-competitive, it suffices to prove
\begin{equation}\label{eq:delay-compressed-goal}
 \sum_j\frac53x_j\le5\OPT^*;
\end{equation}
since together with the two cost bounds above, this inequality gives $\ALGDELAY(\sigma)\le 5\OPT^*+M
\le 5(\OPT(\sigma)-M)+M\le 5\OPT(\sigma)$.

\paragraph{The potential.}
We examine the actions of DMF and $\OPT^*$ in chronological order. When both schedules act at the same time, put all $\OPT^*$ actions before all DMF actions, preserving each schedule's internal order. Thus the retained offline access for group $R_j$ comes before DMF exactly when $\tau_j\le t_j$.

We define our potential as follows: \begin{equation}\label{eq:delay-potential}
 \Phi=\frac53\varphi
      +\sum_{j\in\mathcal P}(5y_j+\mu_j)
      -\sum_{j\in\mathcal L}\frac{10}{3}x_j.
\end{equation} We define its terms below.

\textbf{Inversions $\varphi$}: An inversion is an unordered pair of elements appearing in opposite orders in the DMF and $\OPT^*$ lists ($\OPT^*$ list is the same as $\OPT$ list). Let $\varphi$ count these pairs. 

\textbf{Pre-service credit deposit $5y_j$}:
Let $\mathcal P$ contain the indices $j$ for which $\OPT^*$ has served a request in $R_j$ but DMF has not yet served $R_j$. Call such a group \emph{pre-served}. For each $j\in\mathcal P$, include a deposit of $5y_j$.


\textbf{Account $\mu_j$}: We also keep an account $\mu_j$ for group $R_j$ and element $e_j$ to account for movements that happen while a group is pre-served. It starts at zero at the retained offline access. While the group is pre-served, increase $\mu_j$ by $\frac{10}{3}$ whenever a paid swap by $\OPT^*$ moves $e_j$ one position backward.  

Note that the set $R_j$ contributes in total $5y_j+\mu_j$. When DMF serves $R_j$, we remove both the deposit and the account, so their decrease can offset DMF's cost.

Unlike the preceding upper-bound proofs, in which the online algorithm always acts at deadlines, here we face the possibility that $\ALGDELAY$ may serve the group before $\OPT^*.$ Let $\mathcal L$ contain the indices $j$ for which DMF has served $R_j$ but the retained offline access has not yet occurred.
For each $j\in\mathcal L$, include the negative term $-\frac{10}{3}x_j$, where $x_j$ is the position of $e_j$ in DMF at DMF's service. Adding this term decreases the potential when DMF serves the group. Keep it until the retained offline access and remove it at $\OPT^*$'s access, increasing the potential by $\frac{10}{3}x_j$. 

Observe that the potential may be negative during the execution. We only need its endpoint values: initially it is zero because the lists agree and neither schedule has served any requests; after both finish,
$\mathcal P$ and $\mathcal L$ are empty, so
$\Phi_{\mathrm{final}}=\frac53\varphi\ge0$.

\paragraph{Comparing costs at each action.}
Let $\Delta\mathrm D$ be $\frac53x_j$ at DMF's event $j$ and zero at other actions (since we're excluding $M$ here). Let $\Delta\OPT^*$ be the cost charged to the reduced schedule at the action. We will prove
\begin{equation}\label{eq:delay-event-bound}
 \Delta\mathrm D+\Delta\Phi\le5\,\Delta\OPT^*.
\end{equation}
Summing this inequality will cancel all intermediate potential
values and give~\eqref{eq:delay-compressed-goal}. There are three action types. Observe that, although delay accumulates continuously, it is charged only at service and does not appear in the potential. Releases also change neither the charged cost nor the potential. It therefore suffices to consider the following three action types.

\paragraph{1. A retained offline access to request in $R_j$.}
If DMF has not yet served $R_j$, then $j$ enters $\mathcal P$.
We add $5y_j$ to the potential and start $\mu_j$ at zero. Neither
list changes, and the assigned delay is zero, so
\[
 \Delta\mathrm D+\Delta\Phi=5y_j=5\,\Delta\OPT^*.
\]
If DMF has already served $R_j$, then $j$ leaves $\mathcal L$.
Removing its negative term increases the potential by
$\frac{10}{3}x_j$. Here $\tau_j>t_j$, and therefore
\[
 \Delta\mathrm D+\Delta\Phi
 =\frac{10}{3}x_j
 \le5\left(y_j+\frac23x_j\right)
 =5\,\Delta\OPT^*.
\]

\paragraph{2. A paid offline swap.}
An adjacent swap changes only the swapped pair's inversion status, so the inversion term increases by at most $\frac53$. We must also bound the increase in the $\mu$ accounts.

Only accounts for the backward-moving element can increase. There is at most one: DMF clears all pending requests to that element, so requests in its next group arrive strictly after that service. The old account is therefore removed before the next group's retained offline access can create a new one.

If the backward-moving element has a pre-served group, its account
increases by $\frac{10}{3}$; otherwise no account changes. All other
terms are fixed. Since the swap costs one,
\[
 \Delta\mathrm D+\Delta\Phi
 \le\frac53+\frac{10}{3}=5=5\,\Delta\OPT^*.
\]

\paragraph{3. A DMF service event.}
At event $j$, DMF serves $R_j$ and moves $e_j$ from position $x_j$
to the front. The charge $\Delta\mathrm D$ is $\frac53x_j$, and
$\OPT^*$ pays nothing at this action. We must therefore show that
$\Delta\mathrm D+\Delta\Phi\le0$. Only the inversion term and the
term for group $R_j$ can change. We distinguish whether the retained
offline access for this group has already occurred. 

\emph{Case 1: the retained offline access has not yet occurred.}
There is no deposit or account to remove. Instead, $j$ enters
$\mathcal L$, adding $-\frac{10}{3}x_j$ to the potential.
The move to the front creates at most $x_j-1$ inversions, so
\[
 \Delta\mathrm D+\Delta\Phi
 \le\frac53x_j+\frac53(x_j-1)-\frac{10}{3}x_j
 =-\frac53\le0.
\]

\emph{Case 2: the retained offline access has already occurred.}
Here $j\in\mathcal P$. DMF's service removes the term
$5y_j+\mu_j$ from the potential, where $\mu_j$ is its balance
immediately before service. We will use this decrease to cover the
service charge of $\ALGDELAY$ and the change in inversions.

Let $A_j$ be the elements ahead of $e_j$ in both lists immediately
before this event. Moving $e_j$ to the front creates an inversion
with each element in $A_j$. Each of its other
$x_j-1-|A_j|$ online predecessors is behind $e_j$ offline, so
moving past it removes an inversion. Including the removal of the
group's deposit and account, the full change is
\begin{align}
 \Delta\mathrm D+\Delta\Phi
 &=\frac53x_j+\frac53\Delta\varphi-5y_j-\mu_j\notag\\
 &=\frac53x_j+\frac53\bigl(|A_j|-(x_j-1-|A_j|)\bigr)
       -5y_j-\mu_j\notag\\
 &=\frac{10}{3}|A_j|+\frac53-5y_j-\mu_j.
 \label{eq:delay-preserved-event}
\end{align}
Thus we only need to account for the elements in $A_j$.

To do this, let $F_j$ be the set of elements ahead of $e_j$ at its
retained offline access. This set records the list at that earlier
access and stays fixed; $|F_j|=y_j-1$. Split $A_j$ into two parts:
\begin{itemize}
 \item \emph{Elements already ahead at the offline access.}
 These are the elements of $A_j\cap F_j$, of which there are at
 most $y_j-1$. They are the elements covered by the deposit $5y_j$.

 \item \emph{Elements that became offline predecessors later.}
 Each $f\in A_j\setminus F_j$ was behind $e_j$ at the retained
 offline access and is ahead of it now. Since $\OPT^*$ changes its
 list only through paid adjacent swaps, an intervening swap must
 have moved $e_j$ backward past $f$. The group was pre-served at
 that swap, so $\mu_j$ increased by $\frac{10}{3}$. Different
 elements require different swaps, and the account never decreases.
 Therefore
 \[
  \mu_j\ge\frac{10}{3}|A_j\setminus F_j|.
 \]
 This includes swaps after the retained access at the same time as
 DMF's service: all offline actions at that time precede DMF's
 event in the analysis.
\end{itemize}

Subtracting $\mu_j$ therefore covers all the new predecessors in the second part,
leaving only those already present at the offline access:
\[
 \frac{10}{3}|A_j|-\mu_j
 \le\frac{10}{3}|A_j\cap F_j|
 \le\frac{10}{3}(y_j-1).
\]
Substituting into~\eqref{eq:delay-preserved-event}, the deposit
covers these remaining terms:
\[
 \Delta\mathrm D+\Delta\Phi
 \le\frac{10}{3}(y_j-1)+\frac53-5y_j
 =-\frac53(y_j+1)\le0.
\]

\paragraph{Concluding the cost bound.}
Summing~\eqref{eq:delay-event-bound} over all actions gives
\[
 \sum_j\frac53x_j+\Phi_{\mathrm{final}}-\Phi_{\mathrm{initial}}
 \le5\OPT^*.
\]
Since $\Phi_{\mathrm{initial}}=0$ and $\Phi_{\mathrm{final}}\ge0$,
this proves~\eqref{eq:delay-compressed-goal}.

\paragraph{Delay-aware extension with non-continuous delay functions.}
A jump in our delay function might skip our $\frac{2}{3}x_i$ threshold. We consider a delay-aware version of $\ALGDELAY$ that knows the functions at release, and hence allows service at the time of the jump. On the other hand, left-continuity ensures that the lower delay value still applies. We choose service times that preserve the two delay bounds in \eqref{eq:delay-trigger-properties} when the functions are left-continuous and known in full at release. 

The service actions of the delay-aware $\ALGDELAY$ remain the same; only the rule for choosing their times changes. At a time $t$, let $R_e(t)$ be the pending requests to $e$, whose current position is $x$. For each nonempty $R_e(t)$, compute
\[
 d_e=\inf\left\{u\ge t:
       \sum_{\sigma_i\in R_e(t)}f_i(u)\ge\frac23x\right\}.
\]
The infimum need not belong to the set: at a jump, the delay may still be below the threshold at $d_e$ and exceed it immediately afterward. Service at the real time $d_e$ is allowed and pays the value at $d_e$.

DMF plans to serve an element with the smallest $d_e$, breaking ties by the fixed order. If new requests arrive at or before that time, it processes all releases at that time and recomputes every $d_e$ using the updated pending sets. It also recomputes after each service, using the updated list positions. All computations use only the delay functions of currently pending requests.

\emph{Delay at DMF's service.} We show that, for each element, the total delay of its pending requests never exceeds its current threshold. Pending delay starts at zero. Arrivals add zero delay, and a service clears the served element's pending set without lowering other thresholds. DMF never waits past the smallest $d_e$, where left-continuity keeps the delay at most the threshold. Hence
\[
 \sum_{\sigma_i\in R_j}f_i(t_j)\le\frac23x_j.
\]

\emph{Delay at a later time.} Since each delay function is nondecreasing, the total delay of the fixed group $R_j$ is also nondecreasing. Fix the same group $R_j$ and a time $u>t_j$. If its total delay were still below $\frac23x_j$ at $u$, monotonicity would keep it below this threshold from the last recomputation through $u$. The infimum defining its service time would then be at least $u$, contradicting service at $t_j<u$. Therefore
\[
 \sum_{\sigma_i\in R_j}f_i(u)\ge\frac23x_j
 \qquad(u>t_j).
\]

These are~\eqref{eq:delay-trigger-properties}, so the preceding cost and potential analysis still gives factor $5$. With continuous functions, the rule reduces to the original equality trigger. Every request is eventually served. Otherwise, an unserved request's delay would eventually exceed $2n/3$, contradicting the bound on the total pending delay at its element.
\end{proof}

\paragraph{Why the threshold is $2/3$.}
Suppose the threshold were $\theta x$, with $\theta>0$. With continuous delays, the first access and delay would cost $(1+\theta)x$. Keeping the same potential form replaces $\frac53$ by $1+\theta$ and $\frac{10}{3}$ by $2(1+\theta)$. For a proposed competitive factor $C$, use deposits $Cy_j$. Two offline events determine the bound:
\begin{itemize}
 \item A paid swap costs one and increases the inversion term and one account by at most $(1+\theta)+2(1+\theta)$. Thus it suffices to have $C\ge3(1+\theta)$.
 \item If DMF serves first, the later offline access removes a negative term, increasing the potential by $2(1+\theta)x$. Its assigned delay $\theta x$ covers this increase after multiplication by $C$ if $C\theta x\ge2(1+\theta)x$.
\end{itemize}
Thus the proof gives the bound
\[
 C=\max\left\{3(1+\theta),\frac{2(1+\theta)}{\theta}\right\}.
\]
The first expression increases with $\theta$, while the second, $2+2/\theta$, decreases. Their maximum is smallest where they are equal:
\[
 3(1+\theta)=\frac{2(1+\theta)}{\theta}
 \quad\Longleftrightarrow\quad
 3\theta=2.
\]
Hence $\theta=2/3$, and both expressions equal $5$. This minimizes the bound supplied by this potential analysis.

\paragraph{Nonzero delay at release.}
If $f_i(r_i)>0$ is allowed, run the algorithm on the residual functions $\bar f_i(t)=f_i(t)-f_i(r_i)$. Every schedule pays the unavoidable amount $B=\sum_i f_i(r_i)$. Adding this amount back to the residual bound gives
\[
 \ALGDELAY_f\le B+5\OPT_{\bar f}\le5\OPT_f.
\]
Thus the factor is unchanged when the trigger uses delay accumulated since release.



\label{main-text-end}

\section*{Acknowledgements}
The main ideas in the algorithms and proofs were developed before any use of generative AI. The authors subsequently used OpenAI's ChatGPT and Codex to assist with checking mathematical arguments, conducting computational experiments, tightening some bounds, and revising the exposition and \LaTeX{} source. The authors critically reviewed all AI-assisted material and take full responsibility for the claims and content of the paper.

\bibliographystyle{splncs04}
\bibliography{ref}
\clearpage
\appendix
\renewcommand{\theHsection}{appendix.\Alph{section}}

\section{Omitted Material from Section~\ref{sec:det}}\label{app:sec3}

\subsection{Counterexamples for simple serving orders}
\label{app:simple-orders}

\lemSimpleOrders*

\begin{proof}[Proof of part~\textup{(a)}]
First, we construct the counterexample for any deterministic $\ALG$ that always serves requests at their deadlines. Let the initial list be $(e_1, \dots, e_n)$.

All $n^2$ requests arrive at time $0$, and request $\sigma_i$ has deadline $d_i = i$. Partition the requests into groups $G_1, \cdots, G_{n}$ where $G_g=\{\sigma_{1+(g-1)n}, \sigma_{2+(g-1)n}, \cdots, \sigma_{gn}\}.$ Each group contains one request to every element. In deadline order, the requested elements in $G_g$ are
\[
\begin{cases}
(e_1,e_2,\ldots,e_n), & \text{if } g \text{ is odd},\\
(e_n,e_{n-1},\ldots,e_1), & \text{if } g \text{ is even}.
\end{cases}
\]
Thus, since $\ALG$ serves requests only at their deadlines, it must serve the groups consecutively, alternating between the forward and reverse order of the initial list.

We claim that each group costs $\ALG$ at least $n(n+1)/2$. Each element starting at position $j$ in when we start serving the group would cost at least $j$ to serve and access (free moves of other elements won't bring it forward, and paid exchanges bringing it forward can be charged to this element). The total cost of each group is therefore at least $1+2+\cdots+n = \Omega(n^2),$ and summing over $n$ groups, 
\[
\ALG(\sigma)\ge\frac{n^2(n+1)}2.
\]

$\OPT$ can instead serve all requests at time $0$, grouping them by element. It first serves all requests to $e_1$, then all requests to $e_2$, and so on, moving each element to the front after its first access. Each element has $n$ requests. The first access to $e_j$ costs $j$. Its remaining $n-1$ accesses cost one each. Hence, 
\[
\OPT(\sigma)\le\sum_{j=1}^n(j+n-1)
=\frac{3n^2-n}{2},
\qquad
\frac{\ALG(\sigma)}{\OPT(\sigma)}
\ge\frac{n(n+1)}{3n-1}=\Omega(n).
\]

Similarly, for an algorithm $\ALG$ that serves every request upon arrival, release the requests at times $1,\ldots,n^2$ in the same order and give them all deadline $n^2+1$. The online algorithm again accesses every element once per group, so the same lower bound applies. Offline waits until time $n^2+1$ and serves the requests element by element using the schedule above, with the same cost. Thus the arrival-only rule also has competitive ratio $\Omega(n)$.

\end{proof}

\begin{proof}[Proof of part~\textup{(b)}]
There are $n^2$ requests in total, and we construct $n$ groups $G_1,\ldots,G_n$, each containing one request to every element. All requests in group $G_g$ arrive at time $2g-2$. Let $\bar e_g$ be the last element in $\ALG$'s list at this time. The request to $\bar e_g$ has deadline $2g-1$; the other requests have deadline $2n$.

At time $2g-1$, $\ALG$ serves the entire group: first the requests to the preceding elements of $\bar e_g$, then the request to $\bar e_g$. Each group therefore requires one access to every element. By the counting argument in the previous proof, it costs $\ALG$ $1 + 2 + \dots + n$ per group, and hence in total,
\[
\ALG(\sigma)\ge\frac{n^2(n+1)}2.
\]

$\OPT$ serves only the urgent request of each group at its deadline, without changing its list. These $n$ accesses cost at most $n\cdot n=n^2$. At time $2n$, it serves the remaining requests element by element, in initial list order, moving each element to the front after its first access. The remaining $n^2-n$ requests cost at most $n^2-n+n(n-1)/2$. Thus
\[
\OPT(\sigma)\le\frac{5n^2-3n}{2},
\qquad
\frac{\ALG(\sigma)}{\OPT(\sigma)}
\ge\frac{n(n+1)}{5n-3}=\Omega(n).
\]
\end{proof}

All lower bounds above also hold when an additive term $\beta(n)$ is allowed in the competitive ratio. Fix $n$ and repeat the construction on disjoint time intervals. Each repetition costs the online algorithm $\Omega(n^3)$ and
offline $O(n^2)$, including the cost of restoring its list. Repeating sufficiently many times makes the fixed additive term $\beta(n)$ negligible, so the competitive-ratio lower bound remains $\Omega(n)$.



\subsection{The lower bound for the family of algorithms including $\FER$}\label{app:shielded}

This subsection proves Theorem~\ref{thm:lazy-lb} and
Corollary~\ref{cor:fer-lb}. 

Recall that an algorithm is \emph{$h$-local-lazy MTF} if it uses no paid exchanges, moves every accessed element to the front, and serves requests early only within the first $h$ positions. We construct an input on which no request is ever eligible for early service. 

\noindent\textbf{Construction.}

Fix $1\le h<n/2$ and $1\le q\le n-h$. Relabel the initial list as
$L_0=(e_n,\ldots,e_1)$, so that $e_1$ is initially last. We distinguish its first $h$ elements, its last $q$ elements, and the elements between them.
\[
L_0=\bigl(
\underbrace{e_n,\ldots,e_{n-h+1}}_{\text{first }h\text{ elements}},
\underbrace{e_{n-h},\ldots,e_{q+1}}_{\text{middle}},
\underbrace{e_q,\ldots,e_1}_{\text{last }q\text{ elements}}
\bigr).
\]
The instance $I^{(h)}_{n,q}$ has  a \emph{main} request $M_\ell$ to each element $e_\ell$ ($1\le\ell\le n$), and an \emph{extra} request $E_\ell$ to each of the last $q$ elements $e_\ell$ ($1\le\ell\le q$) of $L_0$. We choose release times to prevent early service: along a deadline execution, each request arrives only when its element is outside the first $h$ positions. 
Formally, the requests are
\[
 M_\ell=(e_\ell,r_\ell,2\ell)
 \quad(1\le\ell\le n),\qquad
 r_\ell=
 \begin{cases}
 0,&\ell\le n-h,\\
 2\bigl(\ell-(n-h)\bigr)+1,&\ell>n-h,
 \end{cases}
\]
and
\[
 E_\ell=(e_\ell,\,2(\ell+h)+1,\,2n+2\ell)
 \qquad(1\le\ell\le q).
\]



All windows are valid, since the main requests released at 0 have positive deadlines, and the other delayed main requests and the extra requests have window length $2(n-h)-1>0$. For each $1\le \ell\le q$, the extra request $E_\ell$ arrives after the main request $M_\ell$'s deadline: $2(\ell+h)+1>2\ell$. Thus there is no self-overlap. Releases occur at zero or odd times and deadlines at distinct even times, so no tie-breaking convention is needed. 

\noindent\textbf{Online cost of $A$.} The next lemma shows that the release times force $A$ to follow the deadline order, accessing the last element each time.
\begin{lemma}\label{lem:forced}
Every $h$-local-lazy MTF algorithm $A$ serves all requests of
$I^{(h)}_{n,q}$ at their deadlines and at position $n$. Hence
\[
 A(I^{(h)}_{n,q})=n(n+q).
\]
\end{lemma}

\begin{proof}
Consider the execution that serves every request at its deadline and moves its element to the front. We will show that, throughout this execution, every alive request is to an element outside the first $h$ positions. Thus $A$ cannot make a first early access and must follow the same execution. 

We first show that all requests served at deadlines have access cost $n$. Note that for $\ell\in[n]$, at time $2\ell$, before serving $M_\ell$, the list is
\begin{equation}\label{eq:list-lb}
 (e_{\ell-1},\ldots,e_1,\ e_n,\ldots,e_\ell),
\end{equation}
so the requested element $e_\ell$ is always last. After $M_1,\cdots, M_n$ are served, the list is restored to $L_0$ at time $2n$. For $\ell\in [q], $ we then serve the extra requests $E_\ell$ at time $2n+2\ell$, repeating the first $q$ accesses with the same pattern where $e_\ell$ is at the end when we serve $E_\ell$.

In this execution, we first note that the element requested by each request is at position at least $h+1$ when that request is released: 
\begin{itemize}
\item For $\ell\le n-h$, $M_\ell$ is released when $e_\ell$ is at position $n-\ell+1\ge h+1$. 
\item For $\ell>n-h$, let $s=\ell-(n-h)$. Just before $M_\ell$'s release at $2s+1$, the first $s$ main accesses have pushed $e_\ell$ from position $n-\ell+1$ to position $n-\ell+1+s=h+1$. 
\item For $\ell\in [q]$, after $e_\ell$ is served at time $2\ell$ and moved to front, the next $h$ main accesses push it to position $h+1$ before $E_\ell$ is released at time $2(\ell+h)+1$. These accesses exist because $\ell+h\le n$.
\end{itemize}
Since there is no self-overlap, no other request to the same element is served between a request's release and its deadline. Moving other elements to the front cannot decrease its position. Thus every requested element remains outside the first $h$ positions from its release until its deadline access.

Because $A$ uses no paid exchanges and moves every accessed element to the front, its first deviation from this execution would have to be an early access. Immediately before that access, the two lists would agree, so the requested element would lie outside the permitted first $h$ positions, a contradiction. Hence $A$ follows this execution and pays $n$ for each of the $n+q$ requests.
\end{proof}

\noindent\textbf{Cheaper offline schedule.} To compare this cost with $\OPT$, we now construct a feasible offline schedule. The schedule moves $e_1, \cdots, e_q$ forward and serves their main requests in an order that prepares for cheaper extra accesses.

\begin{lemma}\label{lem:opt-schedule}
On $I^{(h)}_{n,q}$,
\[
 \OPT(I^{(h)}_{n,q})
 \le \frac{n^2+n+q^2+q}{2}+hq.
\]
\end{lemma}

\begin{proof}
Consider the offline schedule that uses the following three stages.
\begin{enumerate}
\item \textbf{At time $0$, serve the middle requests and move $e_1, \cdots, e_q$ forward.} Serve $M_{n-h},\ldots,M_{q+1}$ without moving their elements. Their access positions are $h+1,\ldots,n-q$. Next use $q(n-q-h)$ paid swaps to move the block $(e_q, \cdots, e_1)$ immediately behind $(e_n, \cdots, e_{n-h+1})$, preserving the former's internal order. The list becomes
\begin{equation}\label{eq:parked}
 (e_n,\ldots,e_{n-h+1},\ e_q,\ldots,e_1,\
 e_{n-h},\ldots,e_{q+1}).
\end{equation}
Serve $M_1$ at position $h+q$ without moving it. We handle $M_1$ separately because its deadline $2$ precedes
the next stage. Total cost in this stage is $\sum_{p=h+1}^{n-q}p+q(n-q-h)+ h+q.$

\item \textbf{At time $3$, arrange $e_1, \cdots, e_q$ for their extra requests.}
Serve requests $M_q,M_{q-1},\ldots,M_2$, moving each accessed element to the front. These requests were released at zero and have deadlines at least $4$, so all these services are feasible. Their access positions are $h+1,\ldots,h+q-1$. The list becomes
\begin{equation}\label{eq:sorted}
 (e_2,\ldots,e_q,\ e_n,\ldots,e_{n-h+1},\
 e_1,\ e_{n-h},\ldots,e_{q+1}).
\end{equation}
Perform these accesses before serving $M_{n-h+1}$, which is also released at time $3$. Total cost in this stage is $h(q-1) + \frac{q(q-1)}2$.

\item \textbf{Serve every remaining request at release, without moving.}
These are $M_{n-h+1},\ldots,M_n$ and all extra requests.
These main requests arrive at times $3, 5, \cdots, 2h+1$, and every extra request arrives later, at time at least $2h+3$.
In \eqref{eq:sorted}, $M_{n-h+u}$ has access position $q+h-u$
for $1\le u\le h$. The position for $E_1$ is $h+q$, and
that for $E_\ell$, $2\le\ell\le q$, is $\ell-1$.
The list stays unchanged, so these positions hold at every
remaining service. At this stage, serving these $h$ main requests costs $hq+\frac{h(h-1)}2$; and serving all extra requests cost $h+\frac{q(q+1)}2$.
\end{enumerate}
The sum of costs over all stages is $(n^2+n+q^2+q)/2+hq$, proving the bound.
\end{proof}

\noindent\textbf{Lower bound of cost ratio.} Combining the two lemmas gives
\begin{equation}{\label{eq: qcontrolledbound}}
    \frac{A(I^{(h)}_{n,q})}{\OPT(I^{(h)}_{n,q})}
 \ge
 \frac{n(n+q)}{\tfrac12(n^2+n+q^2+q)+hq}\ .
\end{equation}
The parameter $q$ controls the tradeoff. Each extra request adds $n$ to the online cost, while the
part of the offline bound that depends on $q$ is
$\frac{q(q+1)}2+hq$. Thus increasing $q$ raises the online
cost linearly but the offline bound quadratically.



\begin{proof}[Proof of Theorem~\ref{thm:lazy-lb}]
To choose $q$, we first maximize the right-hand side of \cref{eq: qcontrolledbound} over real values of $q$. Differentiating with respect to $q$ gives the maximizer
\[
q^*=\sqrt{2n(n-h)}-n.
\]
Since $q$ must be a positive integer, we set
\[
q=\max\{1,\lfloor q^*\rfloor\}.
\] Because $0<q^*<n/2<n-h$, this choice satisfies $1\le q\le n-h$ and $|q-q^*|\le1$.

By the two lemmas, the reciprocal of the actual cost
ratio is at most
\begin{align*}
\frac{\OPT(I^{(h)}_{n,q})}{A(I^{(h)}_{n,q})}
&\le
\frac{\tfrac12(n^2+n+q^2+q)+hq}{n(n+q)}\\
&=\frac{q^*+h}{n}
  +\frac{(q-q^*)^2}{2n(n+q)}+\frac1{2n}\\
&\le\frac{q^*+h+1}{n}.
\end{align*}
The equality uses $(q^*)^2+2nq^*=n^2-2nh$. Taking reciprocals and substituting $q^*$ proves the desired bound.
\end{proof}

Note that for algorithms that permit no early services ($h=0$), the bound with $h=1$ applies, since every such algorithm is also $1$-local-lazy MTF.

\begin{lemma}\label{lem:fer-lazy}
$\FER$ is $1$-local-lazy MTF.
\end{lemma}
\begin{proof}
By definition, $\FER$ uses no paid swaps, moves each accessed element to the front, and serves early only at the head.
\end{proof}

\begin{corollary}\label{cor:fer-lb}
Over all list sizes, the worst-case cost ratio of $\FER$ is at least $1+\sqrt2$, even on inputs without self-overlap.
\end{corollary}
\begin{proof}
By Lemma~\ref{lem:fer-lazy}, apply
Theorem~\ref{thm:lazy-lb} with $h=1$.
For every $n\ge3$, this gives an input with
\[
\frac{\FER(\sigma)}{\OPT(\sigma)}
\ge\frac{1}{\sqrt{2-2/n}-1+2/n}
\ge\frac{1}{\sqrt2-1+2/n}.
\]
For any $0<c<1+\sqrt2$, choose an integer $n\ge3$ with
\[
n>\frac{2}{1/c-(\sqrt2-1)}.
\]
The resulting ratio exceeds $c$.
A nonpositive $c$ fails on every nonempty input.
\end{proof}

\begin{remark}
The construction above establishes a lower bound for guarantees of the form $A(\sigma)\le c\,\OPT(\sigma)$, without an additive term. It does not by itself establish the same lower bound when the guarantee allows
\[
A(\sigma)\le c\,\OPT(\sigma)+\beta(n),
\]
where $\beta(n)$ may depend on the list size but not on the number of requests. Indeed, our instance has only $n+q\le2n$ requests and online cost at most $2n^2$, so an allowance of $\beta(n)=2n^2$ would cover its entire online cost.

To rule out such a guarantee, we would need inputs on a fixed number of elements for which $A(\sigma)-c\,\OPT(\sigma)$ grows without bound. This requires an additional construction or repetition argument. Letting $n$ grow in the present construction is insufficient, since $\beta(n)$ may grow with it.
\end{remark}

\subsection{The $\FER$-specific lower bound}
\label{app:tw-uniform-local}

In this subsection, we prove a finite-$n$ lower bound and use it to prove
Theorem~\ref{thm:tw-uniform-local}.

\begin{lemma}\label{lem:fer-finite-lb}
Fix $n\ge2$. If
$\FER(\sigma)\le c\,\OPT(\sigma)+\beta(n)$
for every input $\sigma$ on an $n$-element list, then
\[
c\ge\frac{3n^2}{n^2+5n-3}.
\]
\end{lemma}
\begin{proof}

\noindent\textbf{Construction.} 
Let the initial list be $(e_n,\ldots,e_1)$.
Requests are released in the repeating order
\[
 e_1,\ldots,e_n,\quad e_1,\ldots,e_n,\quad\ldots,
\]
with release time and deadline
\[
 r_i=2i+3,\qquad d_i=2n+2i
\]
for the $i$th request. We check that each window has length $2n-3\ge1$. Also, requests to the same element do not overlap, since $r_{i+n}=d_i+3$. 

We organize the input into groups of $3n$ requests, each group consisting of three repetitions of requests for $e_1, \cdots, e_n$. We analyze the per group cost for $\FER$ and our offline schedule.


\noindent\textbf{Online cost of $\FER$.} We show that $\FER$ costs $3n^2$ per group. $\FER$ serves requests early only through head clearance at a deadline. We first show that this never happens by induction. 

At the first deadline $d_1=2n+2$, the head is $e_n$, and the first request for $e_n$ arrives at $r_n=2n+3$, so head clearance does nothing and $\FER$ serves request $1$ at its deadline.

Now suppose $\FER$ has served requests $1,\ldots,i$, each at its deadline. The element $\hat e_i$ requested by request $i$ is at the front. The next request for $\hat e_i$, if present, arrives at $r_{i+n}=d_i+3$, whereas the next deadline is $d_{i+1}=d_i+2$. Thus the next deadline arrives before the head $\hat e_i$'s next request. Since $\FER$ makes no moves between deadlines, nothing happens until $\FER$ serves request $i+1$ at its deadline. By induction, every request is served at its deadline.

These accesses visit the elements in reverse initial order, so each requested element is always positioned last in the list. Each repetition restores the initial list. Hence every access costs $n$, and every group of three repetitions costs $3n^2$.

\noindent\textbf{Cheaper offline scheme.}
Consider an offline schedule that uses the first repetition to arrange the list in ascending order, and then serve the requests in the second repetition at deadlines and the requests in the third at release, reducing the costs of access in both repetitions.

More specifically, consider three consecutive repetitions, starting with the reversed list $(e_n,\ldots,e_1)$, and let $D$ be deadline of the request to $e_1$ in the first repetition of the current group. We describe the schedule relative to this deadline. 

\begin{enumerate}
\item \textbf{First repetition: arrange the list using free move to front.}
Observe that the request to $e_n$ in the first repetition of the group arrives at $D+1$, while all other
requests in the first repetition have arrived by $D-1$.

At time $D$, serve in this order
 $e_{n-1},e_{n-2},\ldots,e_1,$
moving each element to the front for free after serving it. These requests all have deadlines at least $D$, so the services are feasible. Their access positions are
$2,3,\ldots,n$, and the resulting list is
 $(e_1,e_2,\ldots,e_n).$
 At time $D+1$, serve the newly released request to $e_n$ at
position $n$, without moving it. The first repetition is now
complete, and the list remains ascending.
The total cost of this repetition is
\[
 \sum_{j=2}^{n}j+n=\frac{n(n+1)}2+n-1.
\]

\item \textbf{Second repetition: serve requests at deadlines.}
Serve each request in the second repetition at its deadline and move its element to
the front. The access to $e_j$ costs $j$: before this access,
all earlier services in the second and third repetitions only rearrange smaller-index elements, which were already ahead of $e_j$. Thus the second repetition costs
\[
 \sum_{j=1}^{n}j=\frac{n(n+1)}2.
\]

\item \textbf{Third repetition: serve requests at release.}
Serve each request in the third repetition as soon as it arrives and move its element
to the front. 

If $e_j$ receives its second service at time $t$, its third service is at $t+3$, by $r_{j+n}=d_j+3$. Immediately after its second service, $e_j$ is at the front.
Between the second and third service of $e_j$, only two elements can move
ahead of it: $e_{j-1}$ through its third service at $t+1$
(if $j>1$), and $e_{j+1}$ through its second service at $t+2$
(if $j<n$). There are no other intervening services, since
services within either of these repetitions are two time units apart.
Therefore the first and last accesses in the third repetition cost
$2$ each, and the other $n-2$ accesses cost $3$ each.
The total is
\[
 2+3(n-2)+2=3n-2.
\]
\end{enumerate}

Finally, adding the costs of the three repetitions gives a cost per group of
\[
 \left(\frac{n(n+1)}2+n-1\right)
 +\frac{n(n+1)}2+(3n-2)
 =n^2+5n-3.
\]

Each element's last access serves its request in the third repetition.
These services occur in order $e_1,\ldots,e_n$, leaving the
list reversed again. The last such service occurs at the last release of the request for $e_n$ in the third repetition at time $D+4n+1$;
the next group's first request arrives at $D+4n+3$.
Thus all requests in these three repetitions are served and
the initial list order of $(e_n, \cdots, e_1)$ is restored before the next group begins arriving.
We can repeat the schedule directly, stopping after any complete group.

Fix $n$ and use arbitrarily many groups of three repetitions.
Each group adds cost
$3n^2$ to $\FER$ and $n^2+5n-3$ to the offline schedule.
The fixed additive allowance $\beta(n)$ becomes negligible
per group, so any competitive guarantee requires $3n^2\le c(n^2+5n-3)$, hence
 $$c\ge\frac{3n^2}{n^2+5n-3}.$$
\end{proof}

\begin{proof}[Proof of Theorem~\ref{thm:tw-uniform-local}]
The upper bound follows from Theorem~\ref{thm:fer-upper}. For the lower
bound, Lemma~\ref{lem:fer-finite-lb} gives
\[
c\ge\frac{3n^2}{n^2+5n-3}.
\]
Letting $n\to\infty$ shows that no uniform competitive constant below
$3$ is possible.
\end{proof}

\end{document}